\documentclass[fleqn,usenatbib]{mnras}

\usepackage{newtxtext,newtxmath}

\usepackage[T1]{fontenc}
\usepackage{ae,aecompl}

\usepackage{graphicx}	
\usepackage{amsmath}	
\usepackage{amssymb}	
\usepackage{lastpage}

\title[Stochastic modeling of star-formation histories]{Stochastic modeling of star-formation histories I: \\ the scatter of the star-forming main sequence}

\author[Caplar \& Tacchella]{
Neven Caplar,$^{1}$\thanks{E-mail: ncaplar@princeton.edu}
Sandro Tacchella,$^{2}$\thanks{E-mail: sandro.tacchella@cfa.harvard.edu}
\\
$^{1}$Department of Astrophysical Sciences, Princeton University, 4 Ivy Ln.,  Princeton, NJ 08544, USA\\
$^{2}$Harvard-Smithsonian Center for Astrophysics, 60 Garden Street, Cambridge, MA 02138, USA\\
}

\date{Accepted XXX. Received YYY; in original form ZZZ}

\pubyear{2019}

\begin{document}
\label{firstpage}
\pagerange{\pageref{firstpage}--\pageref{lastpage}}
\maketitle

\begin{abstract} 
We present a framework for modelling the star-formation histories of galaxies as a stochastic process. We define this stochastic process through a power spectrum density with a functional form of a broken power-law. Star-formation histories are correlated on short timescales, the strength of this correlation described by a power-law slope, $\alpha$, and they decorrelate to resemble white noise over a timescale that is proportional to the timescale of the break in the power spectrum density, $\tau_{\rm break}$. We use this framework to explore the properties of the stochastic process that, we assume, gives rise to the log-normal scatter about the relationship between star-formation rate and stellar mass, the so-called galaxy star-forming main sequence. Specifically, we show how the measurements of the normalisation and width ($\sigma_{\rm MS}$) of the main sequence, measured in several passbands that probe different timescales, give a constraint on the parameters of the underlying power spectrum density. We first derive these results analytically for a simplified case where we model observations by averaging over the recent star-formation history. We then run numerical simulations to find results for more realistic observational cases. As a proof of concept, we use observational estimates of the main sequence scatter at $z\sim0$ and $M_{\star}\approx10^{10}~M_{\odot}$ measured in H$\alpha$, UV+IR and the u-band, and show that combination of these point to $\tau_{\rm break}=178^{+104}_{-66}$ Myr, when assuming $\alpha=2$. This implies that  star-formation histories of galaxies lose ``memory'' of their previous activity on a timescale of $\sim200$ Myr, highlighting the importance of baryonic effects that act over the dynamical timescales of galaxies. 
\end{abstract}



\section{Introduction}

Recent observational studies have produced a wealth of data on galaxy populations at different epochs. This has enabled us to study galaxies in a statistical manner and explore scaling relations between a wide range of galaxy properties, such as luminosity, color, stellar mass, size, velocity dispersion, and central stellar mass density, with cosmic time \citep[e.g.,][]{faber76,kormendy77,djorgovski87,baldry04, van-der-wel14a,mosleh17,barro17}. However, how and why individual galaxies change their physical properties and evolve to create such scaling relations is still unclear, since we observe each individual galaxy only once during its cosmic lifetime. \par

A possible way forward is the use of analytical and numerical models. A large body of theoretical work, including cosmological simulations, has been developed aiming to explain the evolving galaxy properties and constrain physical mechanisms that give rise to the observations \citep[e.g.,][]{vogelsberger14, hopkins14, ceverino14_radfeed, porter14, henriques15, schaye15, pillepich18}. Currently, such models still have relatively limited predictive power because of the inadequate spatial/mass/temporal resolution, producing small galaxy samples and/or the uncertainties associated with the ``sub-grid'' physics \citep[e.g.,][]{somerville15}. \par

Another way forward is based on observations: precise evaluation of the number densities of different types of galaxies, coupled with continuity equations, can be used to construct high-fidelity phenomenological models that explain a large number of observed features in the Universe \citep{peng10_Cont,leitner12, carollo13a,damjanov14,kelson14,caplar15,kelson16, caplar18}. Additionally, galaxy stellar ages can be used to understand how galaxies relate to their precursors at earlier epochs to constrain, for example, the mechanism responsible for the change in morphology \citep{fagioli16,williams17,tacchella17_S1,zahid17,damjanov18}. \par

More general than simple ``ages'' are star-formation histories (SFH), which are one of the critical ingredients needed to understand the evolution of galaxies. Having the full information on how galaxies change their star-formation rate (SFR) as a function of cosmic time would allow us to project the galaxy back in time and understand on which timescales the galaxy grew its mass as a consequence of star formation. Furthermore, SFHs encode information about the variability on short and long timescales that arise from different physical mechanisms. There are several physical processes -- acting on different timescales -- that can increase the SFR, such as gas accretion or mergers, and that can decrease the SFR, such as ram pressure stripping of gas or feedback from black holes and stars.\par

SFHs are related to the scaling relation between the SFRs and stellar masses ($M_{\star}$) of galaxies: observational studies have produced a consensus that since at least redshift 3, the majority of star-forming galaxies have a SFR that is strongly correlated with their mass ($\mathrm{SFR}\propto M_{\star}^{\beta}$, with $\beta\sim1$), creating the so-called star-forming main sequence (MS). It has been established that the MS normalization increases with the look-back time as $(1+z)^{\sim 2.2}$ \citep{brinchmann04, daddi07, elbaz07, noeske07b, whitaker12, speagle14, schreiber15, boogaard18}. This increase of the overall normalization of the MS with lookback time can be understood by the higher dark matter accretion rate onto halos, which leads to higher gas fractions in high-redshift galaxies \citep{lilly13_bathtube, tacconi13, tacchella13, tacchella18, genzel15, rodriguez-puebla16}. \par

The most noticeable feature is that the MS relation at any given redshift shows a rather small scatter of \mbox{$\sigma_{\rm MS}\sim0.2-0.4~\mathrm{dex}$}, depending on the exact definition of star-forming galaxies and the method used to infer SFRs and stellar masses. This rather small scatter has been used to argue that the SFRs of galaxies on the MS are sustained for extended periods of time in a quasi-steady state of gas inflow, gas outflow, and gas consumption \citep{bouche10, daddi10, genzel10, dave12, lilly13_bathtube, dekel14_bathtube, tacchella16_MS, rodriguez-puebla16}. Clearly, the MS scatter is directly related to the variability in the SFH and hence encodes physical mechanisms acting on different timescales. In general terms, \citet[][see also \citealt{munoz15}]{abramson15} highlights that if the MS scatter arises due to short-term fluctuations,  star-forming galaxies with similar mass mostly grew-up together, while if the MS scatter arises due to long-term fluctuations \citep{peng10_Cont,behroozi13b} then star-forming galaxies with similar mass did not grow-up together and key physics lies in the process that diversifies star formation histories \citep{gladders13,kelson14}. \par

Based on cosmological zoom-in simulations (VELA simulations; \citealt{ceverino14_radfeed,zolotov15}), \citet{tacchella16_MS} suggested that star-forming galaxies with $M_{\star}\approx10^9-10^{11}~M_{\odot}$ oscillate about the MS ridgeline on timescales of $\leq2$ Gyr at $z\sim1-4$. The propagation of galaxies upwards toward the upper envelope of the MS is due to gas compaction, triggered by e.g., mergers, counter-rotating streams, and/or violent disk instabilities. The downturn at the upper envelope is due to central gas depletion by peak star formation and outflows while inflow from the shrunken gas disk is suppressed. \citet{rodriguez-puebla16} argue, based on advanced abundance matching, that the MS scatter could be set by the halo mass accretion rate, which itself has a scatter of $\sim0.3$ dex, matching the observed dispersion of the MS. \citet{matthee18_MS} show, based on the EAGLE simulations, that the scatter of the MS at $z\sim0$ originates from a combination of short timescales ($\la1~\mathrm{Gyr}$) that are presumably associated with self-regulation from cooling, star formation and outflows, and long time-scale ($\sim10~\mathrm{Gyr}$) variations related to differences in halo formation times. \citet{torrey18} show, based on the IllustrisTNG simulations, that galaxy offsets from the MS and mass-metallicity relation oscillate over similar time-scales (roughly the dynamical time of the dark matter halo) and are often anticorrelated. They speculate that the SFR and metallicity evolution tracks may become decoupled in galaxy formation models dominated by feedback-driven globally bursty SFHs, which could weaken the FMR residual correlation strength. \par

It is well-known that measurements of star-formation rate in different observed bands carry different information, with shorter wavelengths carrying more information about recent star-formation, and longer wavelengths being more sensitive to star-formation on longer timescales. However, measuring SFHs from the observed spectral energy distribution (SED) of galaxies is still very challenging \citep{conroy13_rev}. State-of-the-art SED fitting codes are nowadays able to fit non-parametric SFHs to the galaxy SEDs and provide constraints of the SFHs on timescales of 100 Myr to a few Gyrs, i.e., on rather long timescales \citep{thomas05,graves08,davies15,pacifici16,iyer17,leja18,carnall18}. Short timescales can in principle be inferred from comparing nebular lines to the ultraviolet (UV) continuum. For instance, \citet{weisz12} and \citet{guo16} inferred a systematic decline and increased scatter in nebular line to FUV ratios toward low-mass systems. They interpreted their data with simple SFH models that included bursts lasting tens of Myr, and with periods of $\sim250$ Myr, finding that these low mass galaxies are rather ``bursty''. However, large uncertainties remain, including variations of the initial mass function (IMF), differential dust attenuation distribution in front of young and old stars, and metallicity effects \citep[e.g.,][]{shivaei18}. \par

All this research implies that galaxies exhibits wide range of SFHs, exhibiting bursts, drops and periods non-changing star-formation. In this work, we propose to model the time-dependence of star formation and the movement of star-forming galaxies around the MS ridgeline as a purely stochastic process. We aim to describe the stochastic behaviour in very general terms and therefore define its properties in the frequency domain through a power spectrum density (PSD) that is modelled as a broken power-law. The high-frequency slope of the PSD determines how quickly the SFR changes on short timescales and is going to be connected with physical drivers of star-formation. The broken power-law form allows for a break in the correlation, i.e., sets up a timescale on which the SFR in a single galaxy loses ``memory'' of previous star-formation. \par 

We note that this timescale can, in principle, be longer than the timescale of the Universe. In that case, the PSD is effectively a pure power-law for all practical purposes and the star-formation in a given galaxy has a ``long-term dependence'', i.e., it is correlated over the cosmic timescales.  This process, with a single power-law slope, is used by \cite{kelson14} to model stellar mass growth in the Universe. He described the SFHs using a so-called ``Hurst parameter'', which is very closely related to the slope of the PSD used in this work. We discuss the connection between describing the stochastic process via the PSD with a broken power law (and its high-frequency slope) and the description using ``Hurst parameter'' more extensively in Appendix~\ref{sec:Hurst}. \par

In this paper we focus on the Myr to a Gyr timescale and exclude the longer timescales in the analysis, which we believe set the overall evolution of the MS. Characterisation of the PSD on these ``short'' scales is critical for a couple of reasons. Firstly, it provides a test for simulations, which aim to match various distributions (e.g., galaxy mass function) over a span of redshifts, but do not necessarily aim to match stochasticity of star-formation with cosmic time. Secondly, the recovered parameters of the underlying stochastic process are connected to the physical drivers of star-formation and can provide valuable insight into the properties of star-formation. In particular, the timescale of the break in the PSD should be connected with the timescale on which star-formation due to internal galaxy properties becomes less important compared to external feeding of the galaxy. For an example in a related field, stochastic modelling of the short-term variability ($\sim$ 1 year) of active galactic nuclei (AGN) is one the main methods to explore physical processes powering AGN \citep{kelly09, macleod10, dexter11, macleod12, simm16, caplar17_ptf}, and is providing serious challenges to standard viscous accretion disk models commonly used to model AGN accretion (see review by \citealt{lawrence18}). Finally, understanding the variability of the SFR on short timescale is crucial for proper SED fitting, as the characteristic timescale on which SFR varies is directly related to the size of parameter confidence intervals \citep{leja18_nonpara}. \par 

In this work we lay the basic groundwork of our approach and conduct a simple analysis of the observations in order to constrain the PSD. We focus our investigation on rather massive star-forming galaxies with $M_{\star}\approx10^{10}-10^{10.5}~M_{\odot}$. We work in this rather narrow mass range because measurement errors are the small,  fraction of galaxies that are quenching is negligible, and result are being less affected by possible variations of dust attenuation with galaxies mass (as elaborated in Section~\ref{sec:Observational}). In a companion paper (Iyer et al. in prep.), we analyse and measure the PSDs for a wide range of numerical (cosmological and zoom-in simulations) and semi-analytical models. We find that PSDs are well described by a broken power-law, where the short timescales (high frequencies) follow $\propto f^{\alpha}$ and above a certain decorrelation timescale, $\tau_{\rm break}$, the PSD is constant. Nearly all of the different theoretical models  produce $\alpha\approx-2$, while having different values for $\tau_{\rm break}$, ranging from $\tau_{\rm break}\sim100~\mathrm{Myr}$ in the FIRE simulations to $\tau_{\rm break}\sim1000~\mathrm{Myr}$ in IllustrisTNG. We show that the PSD is set by physical ingredients in the models and, hence, the PSD contains a wealth of information that has not yet been exploited. In a separate, upcoming paper we will address many issues that complicate the precise determination of the PSD in observations, including possible variations in the IMF and dust attenuation. There, we also wish to use observational constraints to analyse properties of galaxies at higher redshifts, where the proper consideration of these complications is critically important. \par

This paper is structured as follows. In Section \ref{sec:Description} we introduce our model and give a short overview of the properties of stochastic processes. In Section \ref{sec:Finding} we show how we can infer parameters describing movement of galaxies about the MS. We show both analytical and numerical results in the simplified case in which response of each star-formation indicator is given by a step function. In Section \ref{sec:RealResponse} we show predictions for the observational results given the more realistic response functions. In Section \ref{sec:Observational} we use results developed in previous sections and determine, as a proof of concept, parameters describing stochastic processes by using observations of the measured scatter of the MS in the local universe. We conclude in Section \ref{sec:Summary}. In the appendix we expand on several more technical aspects of the paper. \par 

Throughout the paper we use the terms ``scatter of MS'', ``standard deviation of MS'' and ``width of MS'' interchangeably, and denote it by $\sigma_{\rm MS}$. We use the term ``intrinsic'' MS, or ``momentary'' MS to denote width of the MS that would be detected if we could measure the star-formation over the previous 1 Myr. We expand on our choice for 1 Myr in Section \ref{sec:Generating}. We make our codes available at \href{github.com/stacchella/variability_SFHs}{github.com/stacchella/} and \href{github.com/nevencaplar/Main-Sequence-Variability}{github.com/nevencaplar/}. \par

\section{Description of star-formation histories with stochastic processes} \label{sec:Description}

In this section, we introduce our approach to modelling SFHs with stochastic processes. In particular, we start by highlighting and justifying the assumptions we make in our model. We then explain what a stochastic process is and how it can be described mathematically. Finally, we show how we generate SFHs with a stochastic process and discuss their properties such as burstiness. \par

\begin{figure*}
    \includegraphics[width=\textwidth]{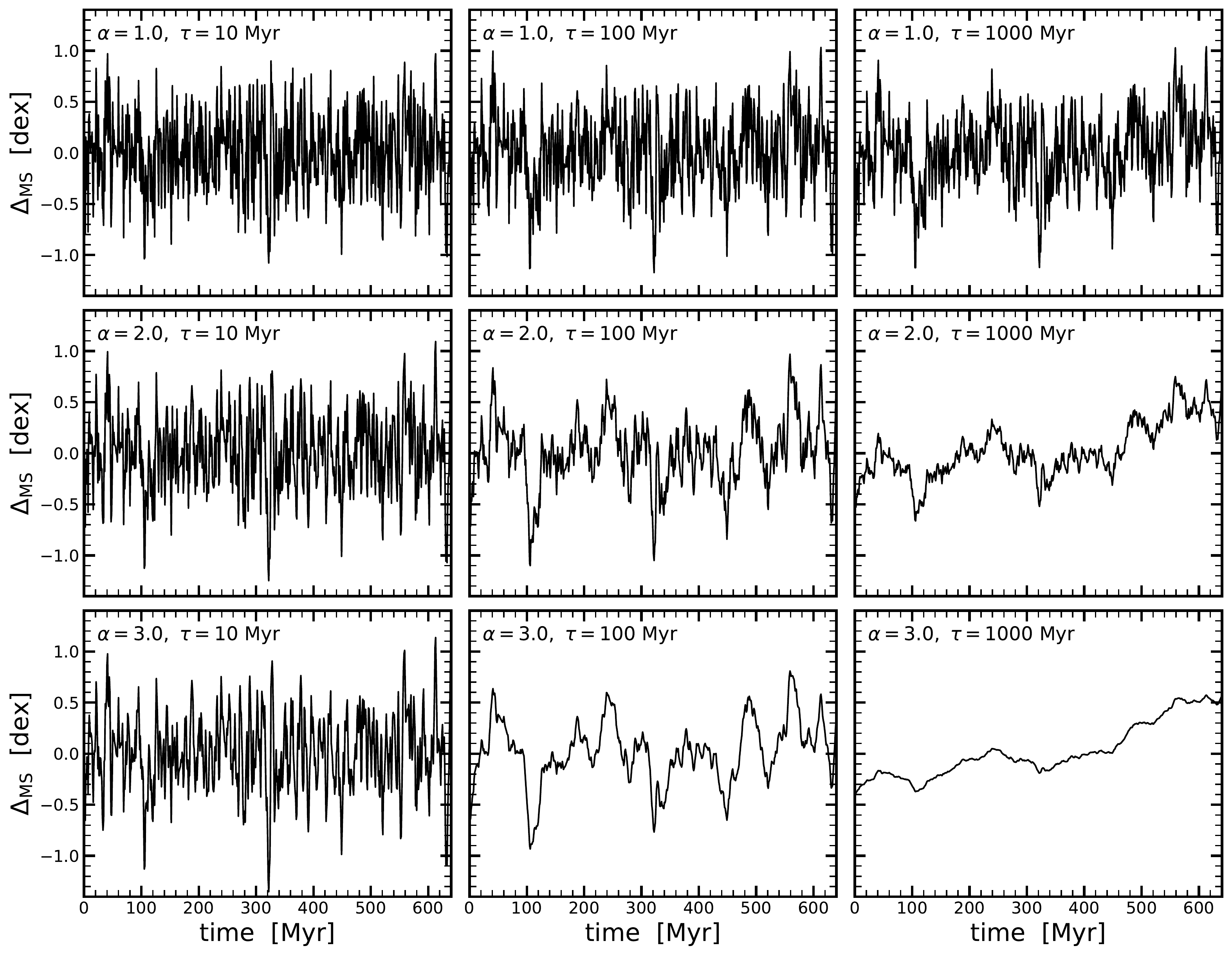}
    \caption{Example star-formation histories (SFHs) drawn from different power spectral densities (PSDs). All of these have been created with the same random seed so they are directly comparable. Each panel shows the SFH relative to the MS ridgeline ($\Delta_{\rm MS}$), measured with cadence (time-separation) of 1 Myr. From top to bottom, we show the SFHs generated with different PSDs with increasing power-law slope ($\alpha=1\rightarrow3$) and increasing break timescale ($\tau_{\rm break}=10\rightarrow1000~\mathrm{Myr}$). Overall, the burstiness increases toward smaller $\alpha$ and shorter $\tau_{\rm break}$.} 
    \label{fig:Example_SFH}
\end{figure*}

\begin{figure*}
    \centering
    \includegraphics[width=0.92\textwidth]{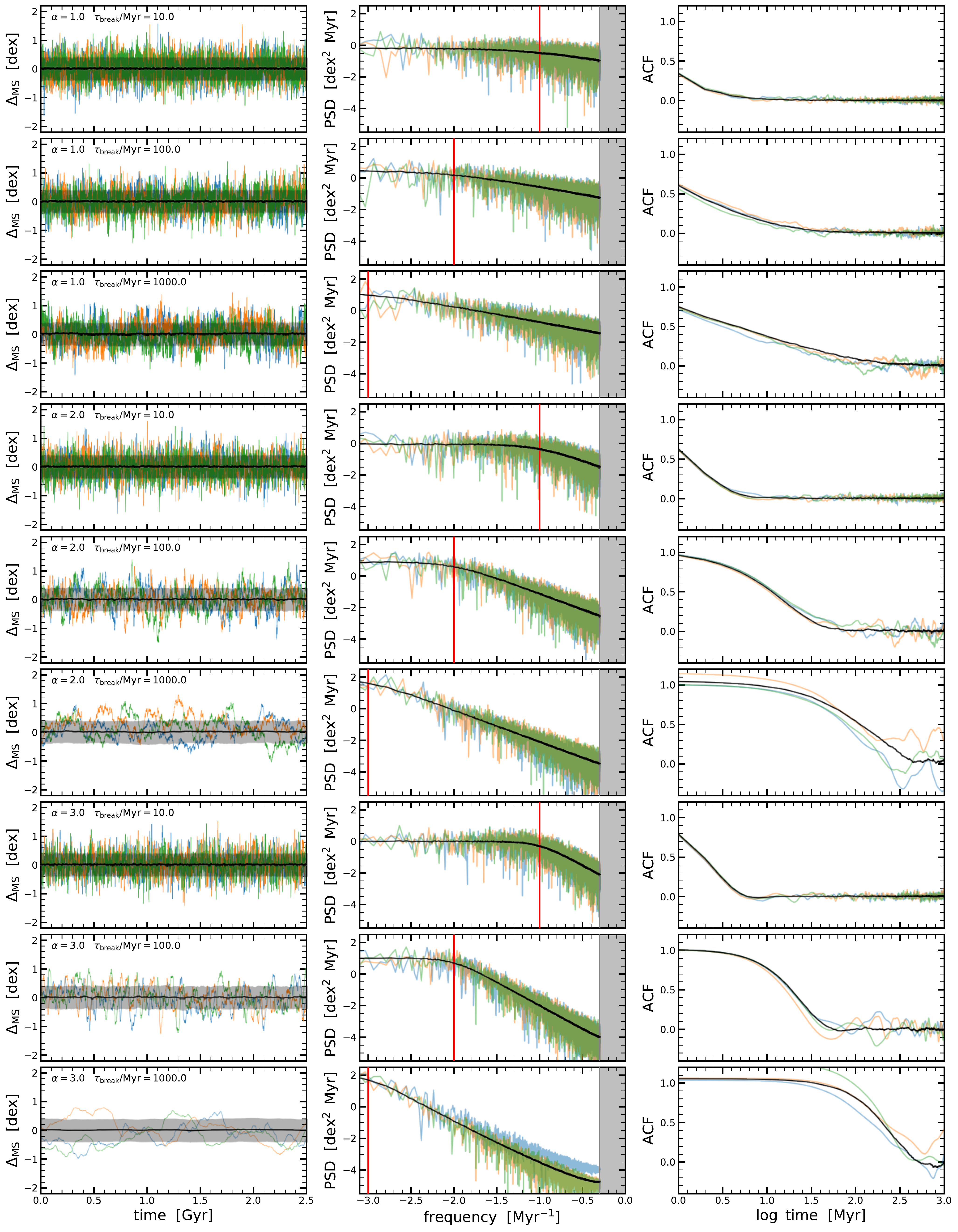}
    \caption{Example SFHs relative to the MS, associated PSDs and ACFs. From top to bottom, we show SFHs with increasing slope ($\alpha=1\rightarrow3$) and increasing break timescale ($\tau_{\rm break}=10\rightarrow1000~\mathrm{Myr}$). The left panels shows the SFHs relative to the MS as a function of time. Three, randomly drawn, examples are shown with blue, orange and green lines. The solid black line and the grey shaded region show the mean and scatter of the MS of our 1000 model galaxies. Given the large number of galaxies in the simulations, the mean relation and the scatter change very little with time. The middle panels shows the associated PSDs; the black line shows the median PSD of our 1000 model galaxies. The vertical red line indicates $\tau_{\rm break}$, while grey region indicates our resolution limit at which we truncate the PSDs. The right hand panels shows the associated ACFs and the black line shows the median ACF for our 1000 model galaxies. } 
    \label{fig:Example_PSD}
\end{figure*}

\begin{figure}
    \centering
    \includegraphics[width=\linewidth]{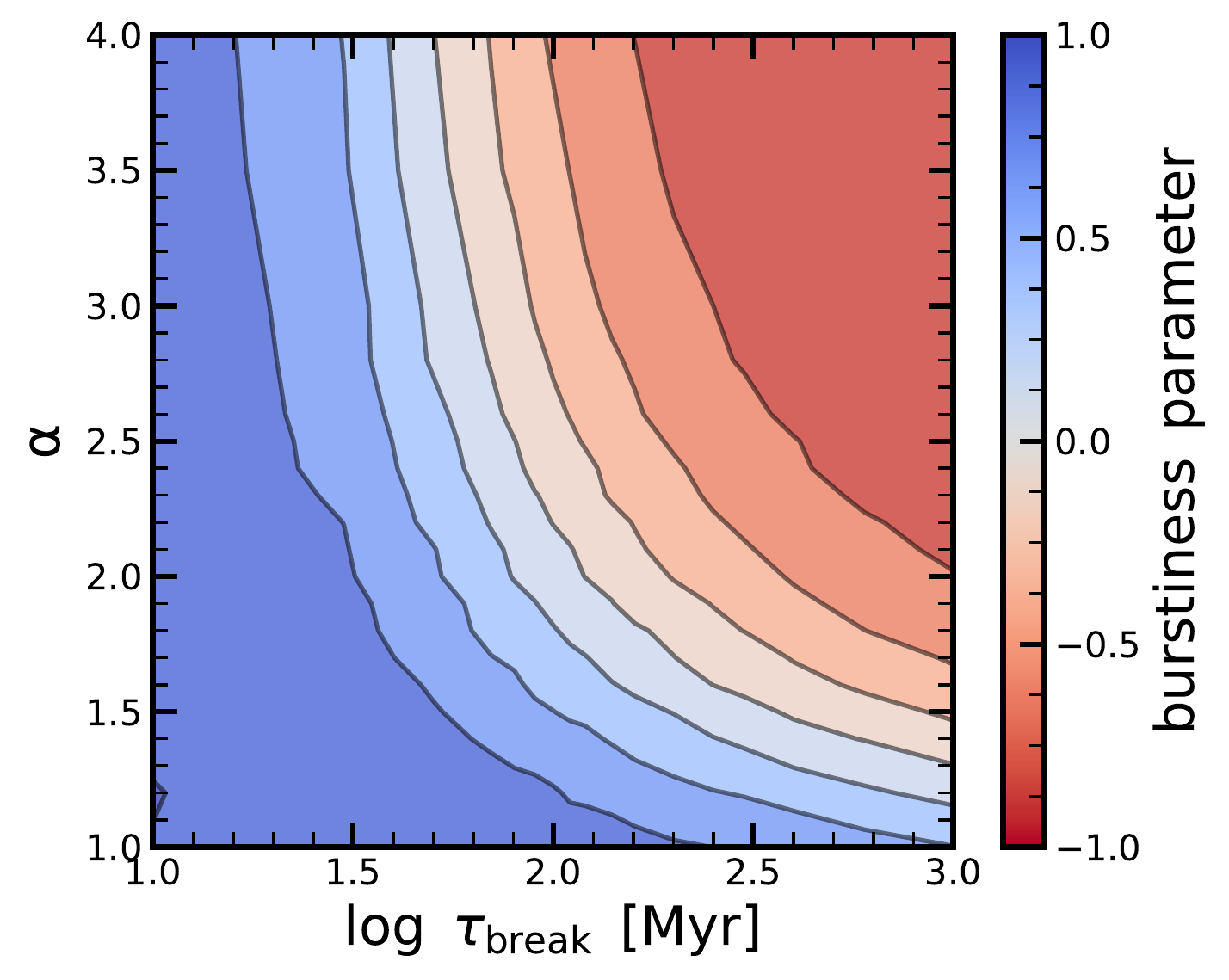}
    \caption{The burstiness parameter $B$. We plot the burstiness parameter (Equation~\ref{eq:burstiness}) estimated for the different PSDs with power-law slope $\alpha$ and break timescale $\tau_{\rm break}$. PSDs with low values of $\alpha$ and $\tau_{\rm break}$ yield highly bursty ($B=1$) SFHs, while PSDs with high values of $\alpha$ and $\tau_{\rm break}$ give constant and hence regular ($B=-1$) SFHs. There is a clear degeneracy between $\alpha$ and $\tau_{\rm break}$.} 
    \label{fig:Burstiness}
\end{figure}

\subsection{Assumptions}

We propose to model time-dependence of the SFR relative to the MS ridgeline of star-forming galaxies as a purely stochastic process. We denote the SFR relative to the MS ridgeline for a galaxy with stellar mass $M_{\star}$ by $\Delta_{\rm MS}=\log(\mathrm{SFR}/\mathrm{SFR}_{\rm MS}(M_{\star}))$, where $\mathrm{SFR}_{\rm MS}(M_{\star})$ is the SFR of the MS at mass of $M_{\star}$. Throughout this work, we assume that:

\begin{itemize}
\item there exists a MS, i.e., there is a power-law relation between SFR and stellar mass for star-forming galaxies that are separate from the quiescent galaxy population;
\item the distribution of SFRs of galaxies about the MS ridgeline is well described by a log-normal function;
\item time-variability of $\Delta_{\rm MS}$ of all star-forming galaxies of a given mass can be described with the same stochastic process given by a broken power-law PSD. 
\end{itemize}

The first two assumptions are well justified from observations. Specifically, the existence of the MS have been put forward by several different observing groups \citep[e.g.,][]{noeske07b, elbaz07, daddi07,peng15}. Furthermore, several groups claim that the scatter can be well described by a log-normal \citep{guo13_MS, schreiber15, chang15, davies19}. \citet{feldmann17_bimodal} shows that the whole galaxy population, i.e., star-forming \textit{and} quiescent galaxies, can be well fitted by a zero-inflated negative binomial distribution, which is only a small modification from the log-normal distribution describing star-forming galaxies, but has the advantage that no division between star-forming and quiescent galaxies is needed. There is still debate in the literature about the exact shape of the MS, including its slope and turnover at high mass at low redshifts, and also about the existence of a MS at high redshifts ($z>3$; \citealt{speagle14, whitaker14, iyer18}). Since we focus in this paper on moderately massive ($\approx10^{10}~M_{\odot}$) galaxies at $z\sim0$ and model the SFRs relative to the MS, this uncertainty on the exact shape is not critical. \par

The third assumption that all massive star-forming galaxies are described with the same stochastic processes is the simplest reasonable assumption and it enables us to maintain analytical tractability of the model. In this work we do not explore further variables that could influence the PSD, such as environment. It is also possible that lower-mass galaxies (including dwarfs) follow a stochastic process with different properties; this will be investigated in the future. Finally, as also mentioned in the introduction, we assume that the stochastic behaviour can be described through a broken power-law PSD, i.e., we assume that the 3-point correlation function (and any higher order terms) is zero. This is again a reasonable assumption given the current observational constraints, as further discussed in Section~\ref{sec:Observational}.\par

\subsection{General description of stochastic processes}

In this section, we will discuss several different stochastic processes and develop an intuition for parameters that we will use in the rest of paper. We characterise different stochastic processes by their PSD, a quantity denoting the amount of power that a process has as a function of frequency, per unit frequency. The PSD gives a quantitative estimate of the importance of each mode, i.e., the importance of each frequency interval for a given stochastic process. We will also extensively use the Fourier pair of the PSD, the autocorrelation function (ACF). The autocorrelation function provides the same information as the PSD, but in the time domain instead of the frequency domain. The definition of ACF, for a process that leads to a time-series $ \Delta(t)$ is:

\begin{equation}
ACF(\delta t)= \frac{\lim_{T -> \infty } \frac{1}{T} \int_{(T)}  \Delta(t) \Delta(t+\delta t) dt}{\sigma^{2}_{ \Delta}}
\end{equation}

\noindent 
where $\sigma^{2}_{ \Delta}$ is the long-term variance of the process. We can see that highly correlated timeseries, where $ \Delta(t) \sim  \Delta(t+\Delta t)$, give ACF$\sim1$, while uncorrelated timeseries will lead to ACF$\sim0$. \par

The simplest stochastic process is white noise. This process has equal power at all frequencies. In this work we denote frequency with $f$. The statement about equal power on all frequencies can be mathematically written as $PSD (f) \propto f^{\alpha}$, with $\alpha=0$. This process is perfectly uncorrelated, i.e., for a white noise process it holds that $ACF(t)=0$, for all times when $t>0$.\par

Other processes can be defined by modifying the slope $\alpha$ of the PSD. By decreasing the slope $\alpha$ the nature of process changes and values separated by a given $\Delta t$ become more similar. Some well known examples are given by $\alpha=-1$ (pink noise) and by $\alpha=-2$ (red noise); the latter corresponding to a random walk or Brownian motion. \par

All natural process have some characteristic time $\tau$, after which a process becomes uncorrelated, i.e., loses its ``memory'' of the previous values and behaves like a white noise process. We can express that statement in terms of the ACF by saying that the ACF at times longer than the decorrelation time, $ \tau_{\rm decor}$ has to be very small, $ACF (t > \tau_{\rm decor}) \ll 1$, or, in the closely related terms of the PSD, by requiring that the slope at small frequencies below ``break'' frequency, $f_{\rm break}$, has to be close to zero, $\alpha(f < f_{\rm break}) \approx 0$. In that case, the PSD has a broken power-law shape, with two distinct slopes at large and small frequencies.\par

To illustrate this point further, let's consider the simplest and most instructive case where the PSD is described by a broken power-law, which corresponds to a damped random walk. The damped random walk is defined by the solution to the stochastic differential equation \citep{kelly15}:

\begin{equation} \label{eq:DRWone}
\frac{d \Delta(t)}{dt}+\frac{1}{\tau_{\rm decor}}  \Delta(t)=\epsilon (t),
\end{equation}

\noindent 
where $\epsilon (t)$ is a continuous white-noise process with zero mean and variance $\sigma_{\rm int}^{2}$. In this form, one can recognise $\sigma_{\rm int}^{2}$ as an intrinsic amount of ``randomness'' entering the system, while $\tau_{\rm decor}$ describes the timescale by which the process gravitates towards a stable-state solution. This process is also known as the Orstein-Uhlenbeck process in physics, or the Vasicek model in financial literature.\par

The damped random walk process leads to the following PSD:

\begin{equation}\begin{split} \label{eq:DRWtwo}
PSD (f) & = \frac{\sigma^{2}_{\rm int}}{ 1/\tau^{2}_{\rm decor}+(2 \pi  f)^{2}}\\
& = \frac{\sigma^{2}}{1+(\tau_{\rm break} f)^{2}}. 
\end{split}\end{equation}

\noindent 
where $\sigma = \sigma_{\rm int} \tau_{\rm decor}$ is  the ``long-term'' variability and $\tau_{\rm break}=2\pi \tau_{\rm decor}$. The connection of the description of the process in frequency and time domain can be achived via the Wiener-Khinchin theorem \citep{emmanoulopoulos10}. Application of this procedure for this particular choice of PSD leads to 

\begin{equation} \label{eq:exampleACF}
ACF(t) = \exp(-t/\tau_{\rm decor}).
\end{equation}

\noindent 
We see that this process has all the properties that we mentioned above. At short timescales (high frequencies), it is very well correlated (ACF $\sim$ 1) and described with a PSD with a slope that equals to 2. Then, the process becomes rapidly decorrelated at a timescale $\tau_{\rm decor}$, and at long timescales (low frequencies), the ACF drops quickly (ACF $\sim$ 0) and the PSD is well described with a flat slope. Observationally, high values of ACF imply similar measurements, while decorrelation implies little to no similarity between two measurements. As such, while at short timescales star-formation measurements will be very similar, at timescales larger than $\tau_{\rm decor}$ the observations of the same galaxy will have star-formation measurements which are largely independent of each other. \par

In this work, we consider a larger group of processes, described with 
\begin{equation} \label{eq:PSD}
PSD (f)  = \frac{\sigma^{2}}{1+(\tau_{\rm break} f)^{\alpha}}.
\end{equation}

\noindent 
This choice offers additional flexibility to encompass a wide range of physical processes that are not necessarily described by random walks, providing a more general description of the possible underlying stochastic process. We show examples of generated SFHs relative to the MS, $\Delta_{\rm MS}$, in Figure \ref{fig:Example_SFH}. The generation of these SFHs is described in the next section. Figure shows simulated SFHs as a function of high-frequency slopes of the PSD ($\alpha$ in Equation \ref{eq:PSD})\footnote{In the rest of the paper $\alpha$ will appear as a positively defined quantity, i.e., $\alpha$ is defined with Equation \eqref{eq:PSD}, and not with $PSD (f) \propto f^{\alpha}$. } and break timescales ($\tau_{\rm break}$ in Equation \ref{eq:PSD}). The figure highlights large qualitative differences between SFHs generated with different parameters. When the slope of the PSD is shallow and/or $\tau_{\rm break}$ is small, the SFH rapidly oscillates around the mean and is very quickly decorrelated. On the other hand, steeper slopes and longer $\tau_{\rm break}$ lead to slower oscillations and more correlated behaviour, as described above. In Section \ref{sec:burst}, we will return to this point in order to quantify the ``burstiness'' of these processes. \par

All of the examples shown in Figure \ref{fig:Example_SFH} have been created with $\sigma=0.4~\mathrm{dex}$. We use this value motivated by the measurement of the MS width in \cite{davies19} in H$\alpha$, and we show in Section \ref{sec:RealResponse} that the measurement in H$\alpha$ is largely representative of the intrinsic width of MS, $\sigma_{\rm MS}$. A number of results that we will derive depend somewhat on the intrinsic width of MS. In the rest of the paper, we work with $\sigma_{\rm MS}=0.4~\mathrm{dex}$, but note at each point where our results depend on the exact choice of this value. \par

Finally, we note that we are implicitly assuming that the stochastic process that we are using is stationary. A stationary process has the property that its mean, variance and autocorrelation function are not time dependent. That the variance, i.e., the width of MS is not time dependent over last several Gyr is consistent with observational data and simulations (\citealt{noeske07b, rodighiero10}; Katsianis et al. in prep.). In other words, we assume that the properties of stochastic process do not change with time and that is valid to use the same parameters to describe observations of galaxies over the whole time range probed. It is important to note that this does not mean that we assume that the MS is not allowed to change with time. The main stochastic process modelled in this work is movement of galaxies \textit{about} the MS ridgeline ($\Delta_{\rm MS}$). The MS itself evolves over cosmological timescales, but because we consider galaxies at $z=0$, this evolution is negligible. We further discuss this effect in Section \ref{sec:Observational}, when comparing the results from our modelling with the observational data.\par

\subsection{Generating star-formation histories} \label{sec:Generating}

Here we describe how we generate numerically a sample of SFHs that we use in the rest of the paper when we analyze properties of stochastic processes and observational consequences. For each choice of break timescale $\tau_{\rm break}$ and high-frequency slope $\alpha$ we simulate 1000 SFHs using the algorithm by \cite{timmer95} and implemented in Python by \cite{connolly15}. For each galaxy, we only consider the SFR as an integrated quantity over the whole galaxy, i.e., we measure only total SFRs over the whole galaxy without deeper considerations of possible difference in different galactic (sub-)components.\par

When simulating SFHs we need to assume some minimal timescale at which to separate our measurements of SFR (time bins). In this work we set this timescale to 1 Myr. This choice is motivated by the free-fall time and the lifetime of giant molecular clouds which are at least of that order \citep{murray10,hollyhead15,freeman17_gmc,grudic18}. Operationally this means that we are assuming that below the scale of 1 Myr, the SFR is perfectly correlated, i.e., remains constant. This is a reasonable assumption since we are considering only total SFRs integrated over the whole galaxy, which itself is made up of several star-forming events (giant molecular clouds). In terms of PSD that means that the domain of the PSD is defined up to scales of $f<1/\mbox{Myr}$. In terms of ACF this leads to a discontinuity in our assumed ACF at 1 Myr, where we set ACF to be equal to unity. The effect of this assumptions is less and less pronounced when the stochastic process has larger $\alpha$ and $\tau_{\rm break}$, since in this case the process is already highly correlated on 1 Myr scales, and our assumption is satisfied by construction. In Section \ref{sec:Observational}, when we compare with observational data at low redshifts, the deduced parameters are outside the range that is critically sensitive to this assumption, but this might not be the case for studies at higher redshifts where deeper investigation of this assumption should be conducted.\par

For each SFH we simulate 4282 points, corresponding to 4.282 Gyr, i.e., to a galaxy evolving from $z=0.4$ to $z=0$. We note that choice of the length is somewhat aribtary and carries no great physical significance. We often use much shorter SFHs and our results are not dependent on the actual duration of the simulated histories, provided they are significantly longer than the $\tau_{\rm break}$ of the considered process. In Figure \ref{fig:Example_PSD} we illustrate the results of this procedure. For each $\tau_{\rm break}$ and $\alpha$, we show a selection of three SFHs relative to the MS (left panel), together with their PSDs and ACFs (middle and right panels). We also show median relations in black, which have been deduced from all 1000 simulated galaxies. We also explicitly show the effect of a cut at 1 Myr, seen as the grey region in the plots of the PSD, and seen in the fact that ACFs extended ``only'' to $\log(\mathrm{time}/\mathrm{Myr})=0$. \par

\subsection{Quantifying the burstiness of star-formation histories} \label{sec:burst}

Here we want to connect the parameters that we have introduced ($\alpha$ and $\tau_{\rm break}$) with a more intuitive quantity which describes the tendency of the SFH to have ``bursts'', i.e., periods of star formation which are much higher (or lower) than the mean of the population. From Figures~\ref{fig:Example_SFH} and \ref{fig:Example_PSD} it is apparent that the power-law slope $\alpha$ and the break timescale $\tau_{\rm break}$ are connected with the variability, or ``burstiness'', of the SFH. Following \citet[][see also \citealt{applebaum18}]{goh08}, we define the burstiness parameter, for a single galaxy, as

\begin{equation}
B = \frac{\sigma/\mu-1}{\sigma/\mu+1},
\label{eq:burstiness}
\end{equation}

\noindent
where $\sigma$ is the standard deviation of $\Delta_{\rm MS}$, and $\mu$ is the mean of $\Delta_{\rm MS}$ of an individual SFH, measured over a certain time-interval. It is important to note that the exact value of $B$ depends on the interval over which it is measured. We choose this time interval to be 30 Myr in order to be able to describe the large diversity of variability in the SFHs seen in Figure~\ref{fig:Example_SFH}. This choice of time interval is also consistent with the one used by \citet{applebaum18}. In general, $B$ has a value in the bounded range $(-1,1)$, and its magnitude correlates with the signal's burstiness. The average deviation from the MS for one galaxy, $\mu$, will approach zero as the SFH becomes more bursty. Therefore, $B=1$ is a maximally bursty signal, $B=0$ is neutral, and $B=-1$ corresponds to a completely regular (periodic or constant) signal. \par

To show trends as a function of $\alpha$ and $\tau_{\rm break}$, we compute the mean of $B$ over our sample of 1000 numerically generated model galaxies for each point of the $\alpha$-$\tau_{\rm break}$ parameter space. Figure~\ref{fig:Burstiness} shows how the burstiness $B$ varies as a function of the PSD parameters $\alpha$ and $\tau_{\rm break}$. Large values of $\alpha$ and $\tau_{\rm break}$ correspond to slowly varying, nearly constant SFHs (see Figure~\ref{fig:Example_SFH}), corresponding to $B=-1$, i.e., highly non-bursty. Most individual galaxies rarely venture to the edges of the MS and therefore experience no episodes of very high (or very low) SFR in comparison to the average over 30 Myr. On the other hand, PSDs with small $\alpha$ and short $\tau_{\rm break}$ produce SFHs with $B=1$ (highly bursty). These SFH rapidly change their SFR and have very varied SFHs over the considered time range of 30 Myr. Furthermore, Figure~\ref{fig:Burstiness} highlights the degeneracy between $\alpha$ and $\tau_{\rm break}$: contours of constant $B$ extend along well-defined curves in the $\alpha$-$\tau_{\rm break}$ plane. This is something that we will encounter again in sections~\ref{sec:Finding} and~\ref{sec:RealResponse}, where we investigate how the scatter of the MS as seen from different indicators depends on the PSD parameters $\alpha$ and $\tau_{\rm break}$. \par

\section{Finding parameters of stochastic star-formation history in a toy model} \label{sec:Finding}

In the previous section we have shown how different choices for parameters describing the PSD (namely the slope $\alpha$ and the break timescale $\tau_{\rm break}$) lead to different SFHs of star-forming galaxies. Estimating these stochastic parameters from observations is difficult, since it is still a big challenge to infer the SFH from a galaxy's SED (see Introduction). By measuring the SFR using different indicators we are able to estimate how the SFR varied in the galaxy's recent past \citep[e.g.,][]{kennicutt98}. Specifically, measurements of the SFR with indicators that are based on nebular lines will probe very recent star-formation ($\sim10^6~\mathrm{yr}$), while measurement in the UV and optical bands will be sensitive to much longer timescales ($\sim10^7~\mathrm{yr}$ and $\sim10^8~\mathrm{yr}$, respectively; see also Section~\ref{sec:RealResponse}). As we will show below, by studying a large sample of galaxies for which the SFRs have been estimated with different indicators, the parameters of the underlying stochastic process can be deduced. \par

To develop understanding, we present the simplest possible case in which measurements with different indicators correspond exactly to averaging the recent SFR over a given timescale. We will develop analytical expressions for the offset as well as scatter of the MS as a function of the averaging timescale. This approach will help us to see connections and correlations between different parameters more clearly. Afterwards, in Section \ref{sec:RealResponse}, we discuss realistic response functions for different SFR indicators and redo much of the calculations with more complex and more realistic models. In contrast with mainly analytical approach in this section, when dealing with more realistic cases we will derive results with greater emphasis on numerical calculations.\par

\subsection{Impact of averaging of the star-formation histories on the shape of the measured MS}

In this section we show how the measured normalization and width of the MS change as a function of the timescale over which the SFH is averaged. These changes depend on the the properties of the stochastic process (i.e., PSD) and, hence, can also be used to infer these properties. We first show analytically derived relations between the measured offset from the MS and then the expression for the width of the MS. We then confirm these results by our numerically generated model galaxies (Section~\ref{sec:Generating}) and show how parameters of stochastic process can be deduced from measurements. \par

\subsubsection{Mean measured offset from the main sequence}

We first consider mean offsets from the MS, as a function of the parameters of the stochastic process, when measured by indicators probing different timescales. We denote the SFR that has been measured with a given indicator by $\mathrm{SFR_{\rm ind}}$. As already mentioned, we assume now the simplest possible case in which the response of each indicator is given by a step function each with different lengths, i.e., different response times. In that case, the response step function can be divided into $N_{j}$ equal time steps. The measurement of the SFR with a given indicator at time $t_{i}$ is then equal to the averaged ``true'' SFR of the previous $N_{j}$ steps, where $t_{j}$ are labels for these equidistant time steps:

\begin{equation} \label{eq:avgSFR}
\mathrm{SFR}_{\rm ind}(t_{i})=\frac{1}{N_{j}} \sum^{N_{j}}_{t_{j}=1}\mathrm{SFR}(t_{i}-t_{j}).
\end{equation}
\noindent
Let's now define the logarithmic distance of the SFR from the ridgeline of the intrinsic MS\footnote{We henceforth use just $\Delta$ as shorthand for $\Delta_{\rm MS}$ in equations, in order to reduce the size and complexity of our expressions.}
\begin{equation} \label{definitionDelta}
\Delta(t_{i}) =\log \mathrm{SFR}(t_{i}) - \log \mathrm{SFR}_{\rm MS}(M_{\star},t_{i}),
\end{equation}
\noindent
and its equivalent quantity, logarithmic distance of measured SFR from the ridgeline of intrinsic MS
\begin{equation} \label{definitionDeltaind}
\Delta_{\rm ind}(t_{i}) =\log \mathrm{SFR}_{\rm ind}(t_{i}) - \log \mathrm{SFR}_{\rm MS}(M_{\star},t_{i}).
\end{equation}
\noindent
In this work we are also assuming that the time and mass dependence of $\log \mathrm{SFR}_{\rm MS}(M_{\star},t)$ can be neglected, i.e., $\log \mathrm{SFR}_{\rm MS}(M_{\star},t) \equiv C$, where C is a constant in units of $M_{\odot}~\mathrm{yr}^{-1}$. This is done to simplify the calculation and is indeed true at low redshifts (see further discussion in Section \ref{sec:Caveats}). \par

By combining Equations \eqref{eq:avgSFR}, \eqref{definitionDelta} and \eqref{definitionDeltaind} we find that
\begin{equation} \begin{split} \label{eq:DeltaInd}
\Delta_{\mathrm{ind}}(t_{i}) & = \log \frac{1}{N_{j}} \sum^{N_{j}}_{t_{j}=1} \mathrm{SFR} (t_{i}-t_{j}) - \log \mathrm{SFR}_{\rm MS} \\
& = \log \frac{1}{N_{j}} \sum^{N_{j}}_{t_{j}=1} 10^{\Delta(t_{i}-t_{j})+\log \mathrm{SFR}_{\rm MS}}- \log \mathrm{SFR}_{\rm MS}  \\
&=\log \frac{1}{N_{j}} \sum^{N_{j}}_{t_{j}=1} 10^{\Delta(t_{i}-t_{j})}.
\end{split} \end{equation}

\noindent
The above expression is true for a single galaxy, but in observations we are looking for a statistical quantity $\langle \Delta_{\rm ind} (t_{i}) \rangle$, where we use $\langle\rangle$ to denote the average value of the measured offset from MS, i.e., to denote that the measurement has been averaged over a sample of observed galaxies.\par

$\langle \Delta_{\rm ind} (t_{i}) \rangle$ can be explicitly derived for the case of damped random walk (equations \eqref{eq:DRWone} and \eqref{eq:DRWtwo}). When the galaxy is at a MS distance of $\Delta(t_j)$ at some time $t_j$, its expected distribution at later time $t_i-t_j$ is given by \citep{kelly09,macleod10}:

\begin{equation}
P [\Delta(t_{i}-t_{j})] = \mathcal{N} \left( \Delta (t_{j}) \exp(-\frac{t_{i}-t_{j}}{4 \tau_{\rm decor}}) , \sigma_{\rm MS} \sqrt{1-\exp(-\frac{t_{i}-t_{j}}{2 \tau_{\rm decor}})  }\right) ,
\end{equation}
\noindent
where with $P$ we denote expected distribution and $\mathcal{N}(\mu,s)$ denotes a normal distribution with mean $\mu$ and standard deviation of $s$.  We are actually interested in the quantity that appears in Equation \eqref{eq:DeltaInd}, i.e., the expectation value of the quantity $10^{\Delta(t_{i}-t_{j})}$. The expectation value, $E$, for a lognormal distribution is given by

\begin{equation}
E[10^{\mathcal{N}(\mu,s)}] =10^{\mu + \frac{s^{2} \ln(10)}{2}}.
\end{equation}

\noindent

Therefore, the expectation value of the process at time $t_{i}-t_{j}$ is given by 

\begin{equation} \label{expectationvalue}
E[10^{\Delta (t_{i}-t_{j})}]= 10^{\Delta (t_{j}) \exp(-\frac{t_{i}-t_{j}}{4 \tau_{\rm decor}}) +\sigma^{2}_{\rm MS} (1-\exp(-\frac{t_{i}-t_{j}}{2 \tau_{\rm decor}}) ) \frac{\ln(10)}{2} }.
\end{equation}
\noindent
Combining together the expressions from equations \eqref{eq:DeltaInd} and \eqref{expectationvalue} we obtain:
\begin{equation} \label{eq:meanDelta}
\langle \Delta_{\rm ind} (t_{i})\rangle=  \log \left( \frac{1}{N_{j}}\sum^{N_{j}}_{t_{j}=1} 10^{\Delta (t_{j}) \exp(-\frac{t_{i}-t_{j}}{4 \tau_{\rm decor}}) +\sigma^{2}_{\rm MS} (1-\exp(-\frac{t_{i}-t_{j}}{2 \tau_{\rm decor}}) ) \frac{\ln(10)}{2} } \right). 
\end{equation}
\noindent
We see that the measured offset depends on the duration of the response of an indicator ($N_{j}$), the properties of underlying stochastic process (only $\tau_{\rm decor}$ in case of the damped random walk) and the intrinsic width of the main sequence ($\sigma_{\rm MS}$). We can also see that larger offsets are realized for larger values of $N_{j}$, $\tau_{\rm decor}$ and $\sigma_{\rm MS}$. We will investigate the interplay of these parameters more closely in Section \ref{sec:Comparisontonumerically}.\par

\subsubsection{Width of the MS} \label{sec:widthofMS}

We now turn to the width of the MS and explore this property as a function of parameters of the stochastic process. The derivation is somewhat involved; see Appendix \ref{sec:ConnectionAnal} for details. There we show that the lowest order expansion, as a function of  $\sigma_{\rm MS}$, of the measured variance of a sample of galaxies, $\sigma^{2}_{\rm ind,MS}$, is given by

\begin{equation} \label{eq:VarEquation}
\sigma^{2}_{\rm ind, MS} (t_{i})=\frac{\sigma^{2}_{\rm MS}}{N^{2}_{j}} \sum_{t_{j}=1}^{N_{j}}\sum_{t_{j^{'}}=1}^{N_{j}} ACF (t_{j}-t_{j^{'}}),
\end{equation}

\noindent
where $\sigma_{\rm MS}$ is the intrinsic width of the MS. This double summation can be simplified, by using symmetry properties of the auto-correlation function, to yield the measured $\sigma_{\rm ind,MS}$ \citep{PriestlyBook,IvezicBook}:

\begin{equation} \label{eq:sigmaACF}
\sigma_{\rm ind, MS} (t_{i})=\frac{\sigma_{\rm MS}}{\sqrt{N_{j}}} \left(1 + 2 \sum^{N_{j}}_{t_{j}=1} \left(1-\frac{t_{j}}{N_{j}} \right)ACF(t_{j})  \right)^{1/2} .
\end{equation}
\noindent
We note that this equation has the same form as the equation for the uncertainty of the mean for the dataset consisting of $N_{j}$ measurements with a homoscedastic measurement error of $\sigma_{\rm MS}$. This is another intuitive way to think about this quantity that we are measuring. In essence, we are exploiting the ergodic property of our sample. Measuring the width of the distribution created by galaxies which are moving stochastically around the MS with an indicator over $N_{j}$ times is equivalent to measuring the mean of a sample of galaxies on MS at $N_{j}$ different times! \par

We see that in case of $ACF(t_{j})=0$, the expression in the brackets of Equation \eqref{eq:sigmaACF} is unity, and therefore the measured scatter of the MS drops very quickly as $\sigma_{\rm MS}/\sqrt{N_{j}}$, which is the familiar expression for the uncertainty of the mean from any analysis that uses independent measurements. On the other hand, in case of highly correlated data ($ACF(t_{j}) \approx 1$), the measured scatter of the MS is essentially unchanged and corresponds to the initial $\sigma_{\rm MS}$. \par

\subsection{Comparison to numerically generated model galaxies} \label{sec:Comparisontonumerically}

\begin{figure*}
    \includegraphics[width=\textwidth]{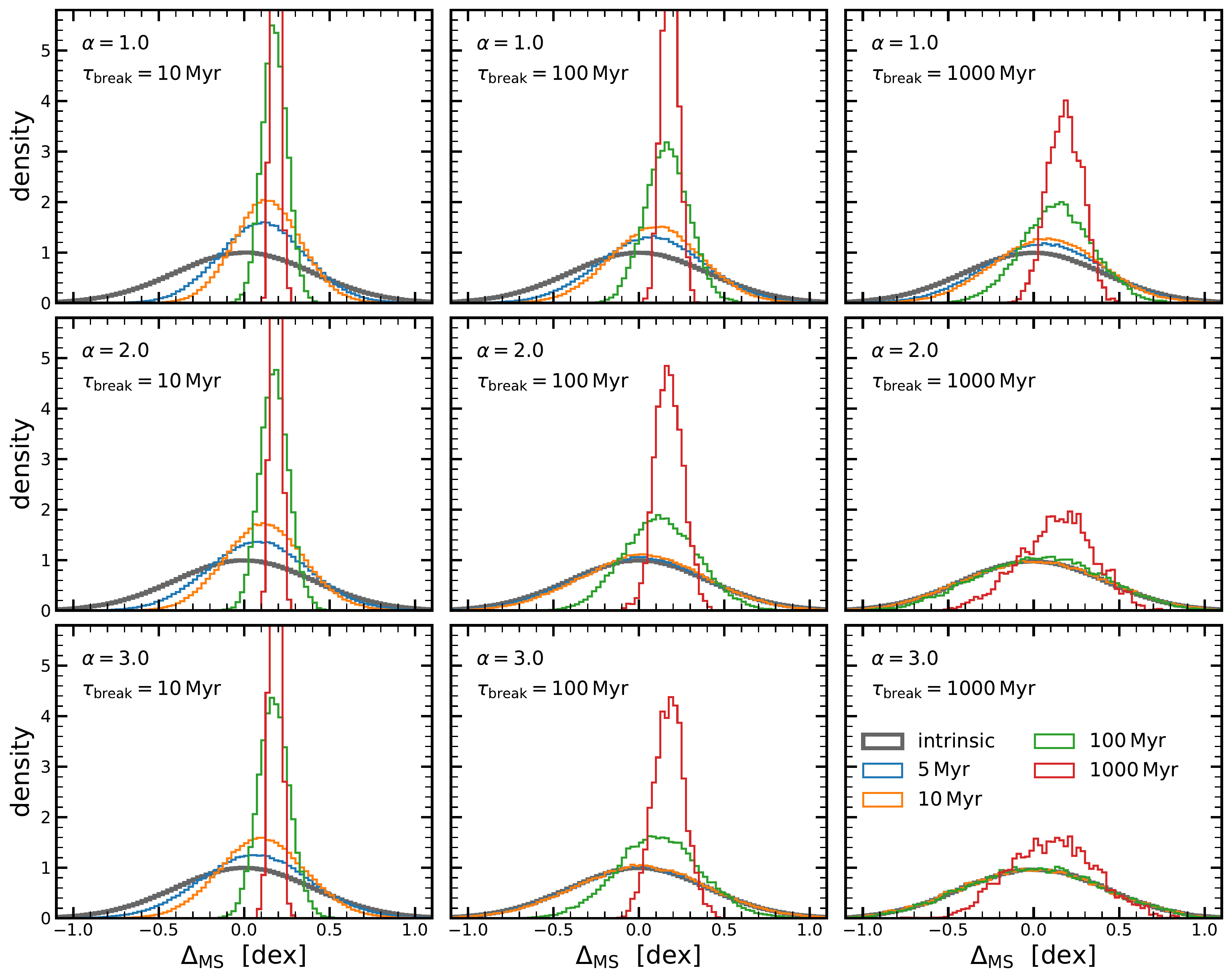}
    \caption{Distribution about the MS ridgeline after averaging the SFHs. Each panel shows the resulting distributions for a given PSD, with increasing $\alpha$ and $\tau_{\rm break}$ from top left to bottom right. The thick black histogram indicates the intrinsic distribution, which has by construction a mean of 0.0 and a scatter of 0.4 dex for all $\alpha$ and $\tau_{\rm break}$. The blue, orange, green and red histograms show the distribution of galaxies after their SFHs were averaged over a timescale of 5 Myr, 10 Myr, 100 Myr, and 1000 Myr, respectively. The larger the averaging timescale, the larger the MS offset and the smaller the scatter. The effects are the largest for small $\alpha$ and $\tau_{\rm break}$ since the SFHs are bursty in this case. }
    \label{fig:DMS_distribution}
\end{figure*}

\begin{figure*}
    \includegraphics[width=\textwidth]{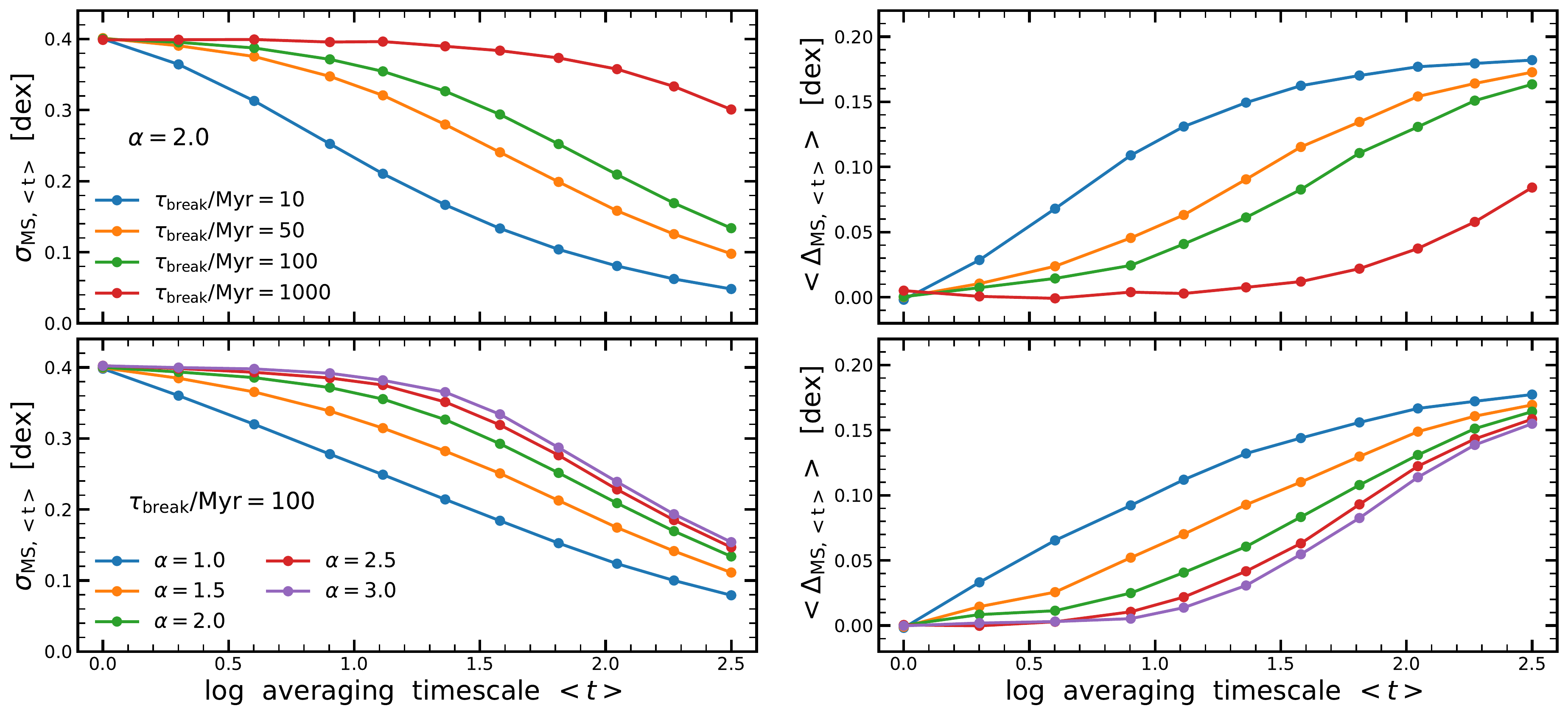}
    \caption{Measured scatter of the MS $\sigma_{\rm MS, \langle t\rangle}$ (left panels) and offset from the intrinsic ridgeline $\Delta_{\rm MS, \langle t\rangle}$ (right panels) as a function of the averaging timescale $\langle t\rangle$. Upper panels show the results for a power-law slope $\alpha=2.0$ and varying $\tau_{\rm break}$, while the lower panels show the results for a constant $\tau_{\rm break}=100.0~\mathrm{Myr}$ and varying $\alpha$. Averaging the SFHs of the model galaxies over longer timescales increases the offset from the MS to higher values and leads to a decrease of the MS scatter, as already seen in Figure~\ref{fig:DMS_distribution}.}
    \label{fig:offset_tavg}
\end{figure*}

We now consider the consequences of the theory developed above, i.e., we study the interplay between the stochastic process of star-formation and the timescale of the indicators used to probe the SFR. We show in Figure~\ref{fig:DMS_distribution}, as a function of PSD parameters, how the shape of the MS (i.e., distribution of $\Delta_{\rm MS}$ at fixed mass) depends on the different averaging timescale. The grey line, which is the same in all panels, shows the intrinsic distribution of the 1000 model galaxies about the MS ridgeline ($\Delta_{\rm MS}$). By construction, this intrinsic distribution is defined in 1 Myr time bins and is a normal distribution with $\sigma_{\rm MS}=0.4~\mathrm{dex}$. \par

Let us first consider the effect of averaging the SFH, i.e., mimicking an observation by averaging the SFR over a certain time interval $t$ (which we denote with $\langle t \rangle$). Note that this is fundamentally different than considering an average of SFH relative to the MS, i.e., considering an average of $\Delta_{\rm MS}(t)$ (see Equation~\eqref{eq:avgSFR}). The blue, orange, green, and red histograms show the distribution about the MS ridgeline after averaging over 5 Myr, 10 Myr, 100 Myr, and 1000 Myr, respectively. As expected, we find that measurements over the shorter averaging timescales give measured distributions that are closer to the intrinsic distribution. On the other hand, long averaging timescales produce a narrower distribution with a median value that is biased to higher values than in the original distribution. This is a consequence of arithmetic averaging of logarithmic quantity (see Equation \eqref{eq:meanDelta}) and serves as a useful reminder that the \textit{measured} median of the MS is not necessarily at the same value as the \textit{intrinsic} median of the MS. \par

We now consider different processes measured with the same indicator, i.e., consider the differences between histograms in different panels with the same colour. We find that the observed distribution matches the intrinsic distribution better for larger values of $\alpha$ and $\tau_{\rm break}$. This is because the SFR of individual galaxies changes less during the time covered by an indicator and is therefore more similar to its momentary value.\par

Therefore, we see that there is an interplay between the parameters of a stochastic process ($\alpha$ and $\tau_{\rm break}$) and the averaging timescale (timescale of the observational indicator). In particular, for a given slope $\alpha$, there is a strong connection between between the timescale of the process and the averaging timescale. This can be seen in Figure \ref{fig:DMS_distribution} as a similarity of the MS distributions that have the same $\tau_{\rm break}$ and averaging timescale in a given row of the figure (compare orange histograms in the panels on the left hand side, green in the middle and red on the right hand side). \par

We note that our formalism is able to predict further moments of the distribution, such as skewness and kurtosis. However, given that those increasingly depend on the details on the assumed intrinsic distribution and are less readily observed quantities, we choose not to extend that part of our analysis.\par

We present the information about offset ($\Delta_{\rm MS, \langle t\rangle}$) and width ($\sigma_{\rm MS, \langle t\rangle})$ of the MS contained in Figure~\ref{fig:DMS_distribution} in a more condensed form in Figure~\ref{fig:offset_tavg}. In the panels on the left we show changes of the measured width of the MS as a function of (a) a fixed slope $\alpha$ and changing $\tau_{\rm break}$, and (b) a fixed $\tau_{\rm break}$ and changing slope $\alpha$. On the right hands side we show the result of the same exercise on the offset of the measured MS ridgeline from the intrinsic MS ridgeline. \par

These plots confirm our intuitive findings from above. For a given indicator corresponding to a single averaging timescale $ \langle t \rangle $, the larger $\alpha$ and shorter $\tau_{\rm break}$ leads to measuring a tighter MS and a larger offset from the median of the intrinsic MS.  \par

\begin{figure*}
    \includegraphics[width=\textwidth]{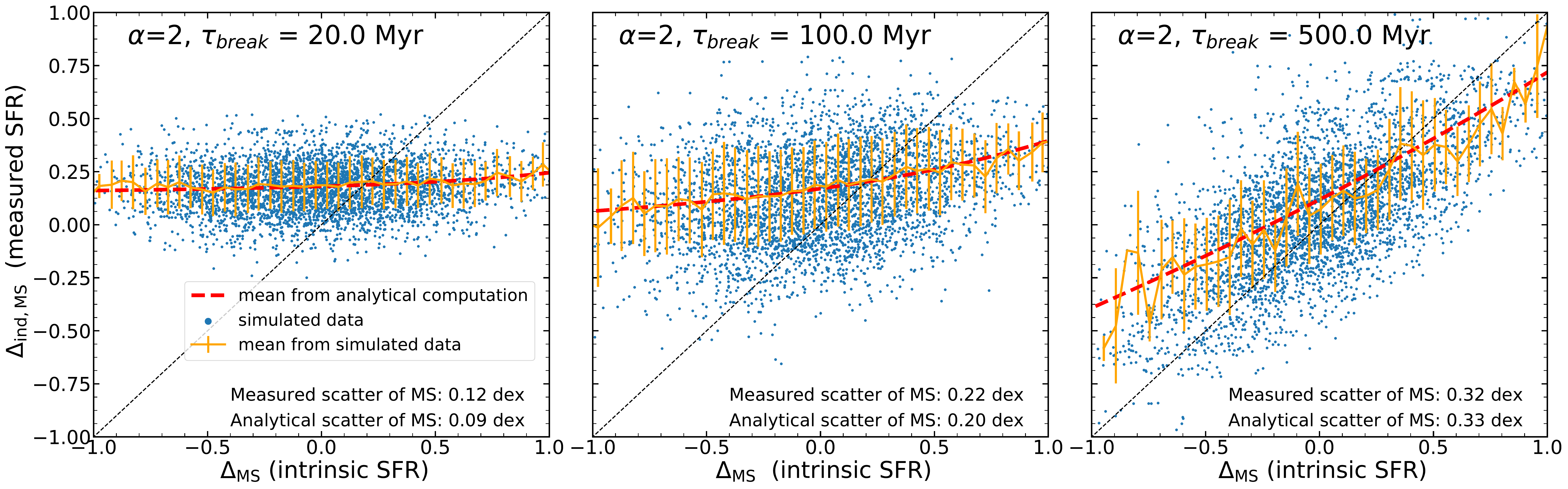}
    \caption{Distribution of measured offset, from the main sequence as a function of intrinsic offset from the main sequence. ``Measurement'' has been approximated in this simple example by averaging previous 100 Myr of SFH of each galaxy.  From left to right, we show result for damped random walk with $\tau_{\rm break}=$20, 100 and 500 Myr. Dashed diagonal line shows one-to-one relation. The orange line shows the mean $\Delta_{\rm ind,MS}$ in the measured data, as a function of input $\Delta_{\rm MS}$, along with the 1-$\sigma$ spread. The dashed red line shows our prediction for the mean $\Delta_{\rm ind,MS}$  in the measured data, calculated from Equation \eqref{eq:meanDelta}. In the bottom right corner we state measured scatter of the MS as seen in the simulated data, and from our analytical expression in Equation \eqref{eq:sigmaACF}. The small offset of the recovered parameters compared to input parameters is due to the fact that Equation \eqref{eq:sigmaACF} is only the first order approximation of the full expression.  } 
    \label{fig:distribution_single}
\end{figure*}

Until now we have discussed how measurements (averaging in this simple toy case) and parameters of stochastic processes influence the statistical parameters of the final distribution such as median and standard deviation. In Figure~\ref{fig:distribution_single} we show a full example of the effect of averaging the SFR (i.e., measuring the SFR with a given indicator) on the observed MS. For all three panels we assume an averaging timescale of 100 Myr and the slope $\alpha=2$, while varying the $\tau_{\rm break}$ of the process. The blue points show the individual model galaxies. The dashed red line indicates the predicted connection between the measured and actual (intrinsic) offset from the MS according to Equation~\eqref{eq:meanDelta}. This prediction is in excellent agreement with the relation measured from simulated model galaxies, which is shown in orange. In the bottom right corner of each panel we also show the measured scatter of the MS and the predicted value from Equation \eqref{eq:sigmaACF}.  \par

In the left panel of Figure~\ref{fig:distribution_single}, we can now explicitly see the central limit theorem in action: because the SFRs vary rapidly, the measured and intrinsic SFRs are almost completely unrelated. This is a consequence of the fact that the observational timescale is much longer than the decorrelation time of the process. This means that during the timescale of the observation, each galaxy has moved up and down the MS ridgeline and hence the measured SFR gives us only an estimate of the mean of the whole population. In the middle and right panels of Figure~\ref{fig:distribution_single}, we increase $\tau_{\rm break}$: the decorrelation times become longer and the agreement between the intrinsic and measured SFRs improve as galaxies change their SFRs less significantly during the time that measurement is sensitive to. \par

\subsection{Measuring parameters of stochastic process from the mock data}

\begin{figure}
    \includegraphics[width=0.47 \textwidth]{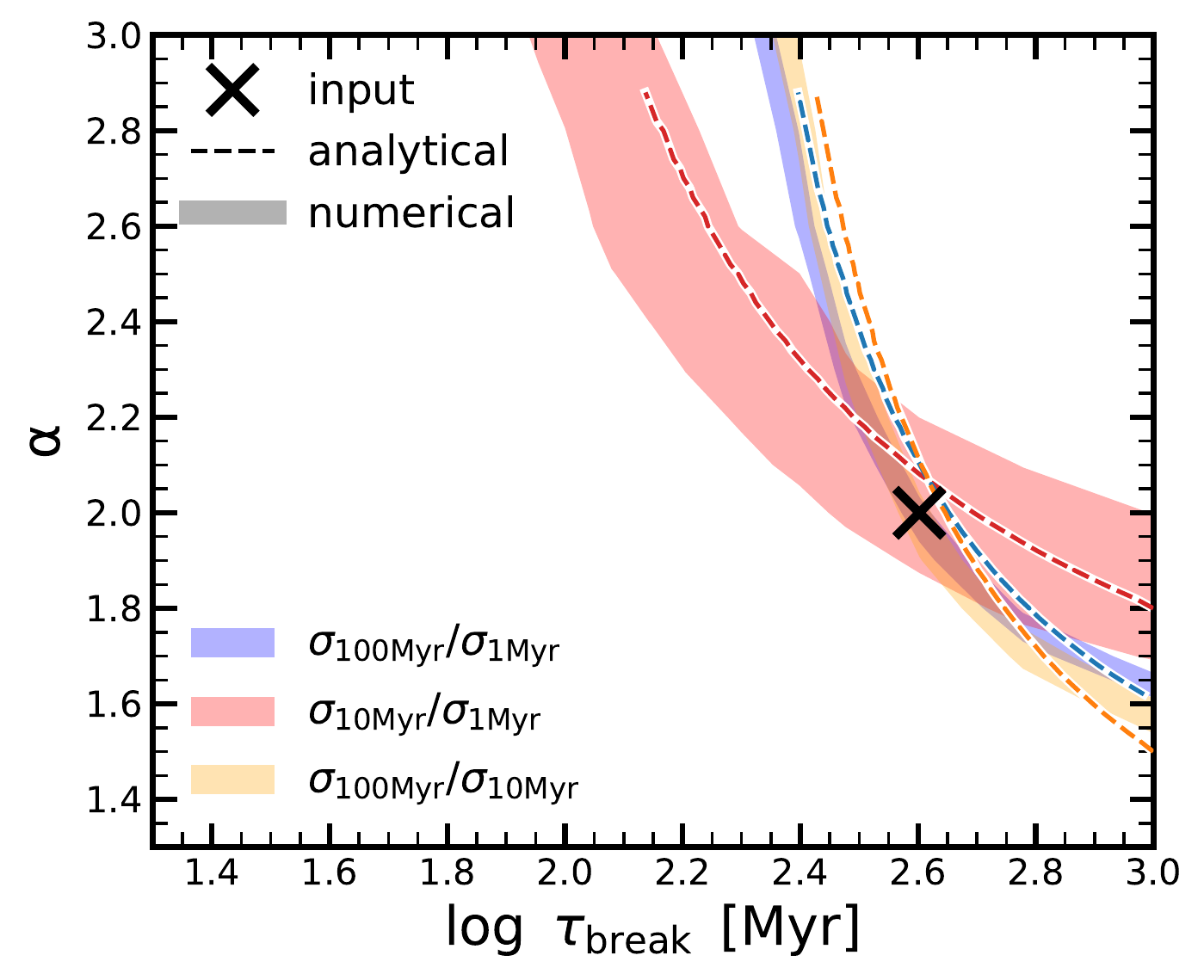}
    \caption{Example of recovering the parameters of a stochastic process by the MS width measurements. The black cross marks the input parameter for the simulated model galaxies ($\alpha=2.0$ and $\tau_{\rm break}=400~\mathrm{Myr}$). The dashed lines show the analytical constraints from the ratios of MS width measurements, while the shaded bands show the numerical constraints. The three different colors show the constraints for three different MS width measurements: a short, medium, and long timescale indicator (1 Myr, 10 Myr and 100 Myr). We find that we can recover the input parameters. The small offset of the analytically recovered parameters compared to input parameters is due to the fact that Equation \eqref{eq:sigmaACF} is only the first order approximation of the full expression. }
    \label{fig:not_degenerate}
\end{figure}

In the previous section we have extensively discussed how averaging, i.e., measuring, SFRs changes the inferred properties of the MS, such as position of ridgeline and width. Furthermore, we have shown how this effect depends on the parameters of the stochastic process itself. Here we want to show how we can use these insights to deduce the parameters of the stochastic process from mock observations. We will then apply the same procedure, with more realistic response functions of different indicators, when discussing real data in Section \ref{sec:Observational}.\par

Here we consider an example in which we have simulated 10000 galaxies with $\alpha=2$ and $\tau_{\rm break}=400$ Myr. We have then convolved these galaxies with three step functions of 1, 10 and 100 Myr respectively, mimicking 3 different indicators. Given that we 1 Myr is the shortest time-scale that we consider, ``measurment'' with the short indicator is the equivalent to the intrinsic SFH. We then measure the ratio of the widths of MS measured in each of these indicators, with associated error bars, and then find parts of the parameter space that are consistent with these measurements, both from considering Equation \eqref{eq:sigmaACF} and from comparing with results from a simulation which spans the whole parameters space. We consider ratios of the measurements in different indicators to minimise the dependence of the result on the intrinsic width of the MS, which is of course known in the simulation, but is not necessarily available when considering real data. We see from Equation \eqref{eq:sigmaACF}, which is first order approximation to the actual measured width, that considering ratios of measured widths eliminates dependence on intrinsic $\sigma_{\rm MS}$. Some dependence comes from higher order terms to Equation \eqref{eq:sigmaACF}, which we discuss in Appendix \ref{sec:ConnectionAnal}. \par 

We show the result of this procedure in Figure \ref{fig:not_degenerate}. The dashed lines show the result deduced from analytic expression in Equation \eqref{eq:sigmaACF}  while the differently coloured areas show 1-$\sigma$ constraints from the numerical simulations. We see that, although a single ratio of MS width measurements of two indicators is not sufficient to break the degeneracy between $\alpha$ and $\tau_{\rm break}$, combination of several indicators can break this degeneracy and provide a unique solution for the parameters of the process. This is not surprising, and is a restatement of the fact that we need two different ``measurements'' in order to uniquely determine two free parameters. It is also advantageous if the indicators cover substantially different range of timescales, in order to be sensitive to different parts of the PSD. Given that we aim to be as agnostic as possible to the intrinsic width of the MS and therefore consider ratios of the MS width, this effectively requires a measurement of the width of the MS with three indicators.  Addition of any more indicators or combinations of indicators will not bring more information in this plane, but can be used as a consistency check. \par

Furthermore, we see that there is a small offset from the input values and the values deduced from using Equation \eqref{eq:sigmaACF}. This is a consequence of the fact that Equation \eqref{eq:sigmaACF} is only the first order approximation to the full calculation of the width, as already discussed in Section \ref{sec:widthofMS} and Appendix \ref{sec:ConnectionAnal}. We use full numerical simulations to avoid this effect here and when we compare with the real data in Section \ref{sec:Observational}. \par

\section{Finding parameters of stochastic star-formation history with realistic response functions}  \label{sec:RealResponse}

\begin{figure*}
    \centering
    \includegraphics[width=\textwidth]{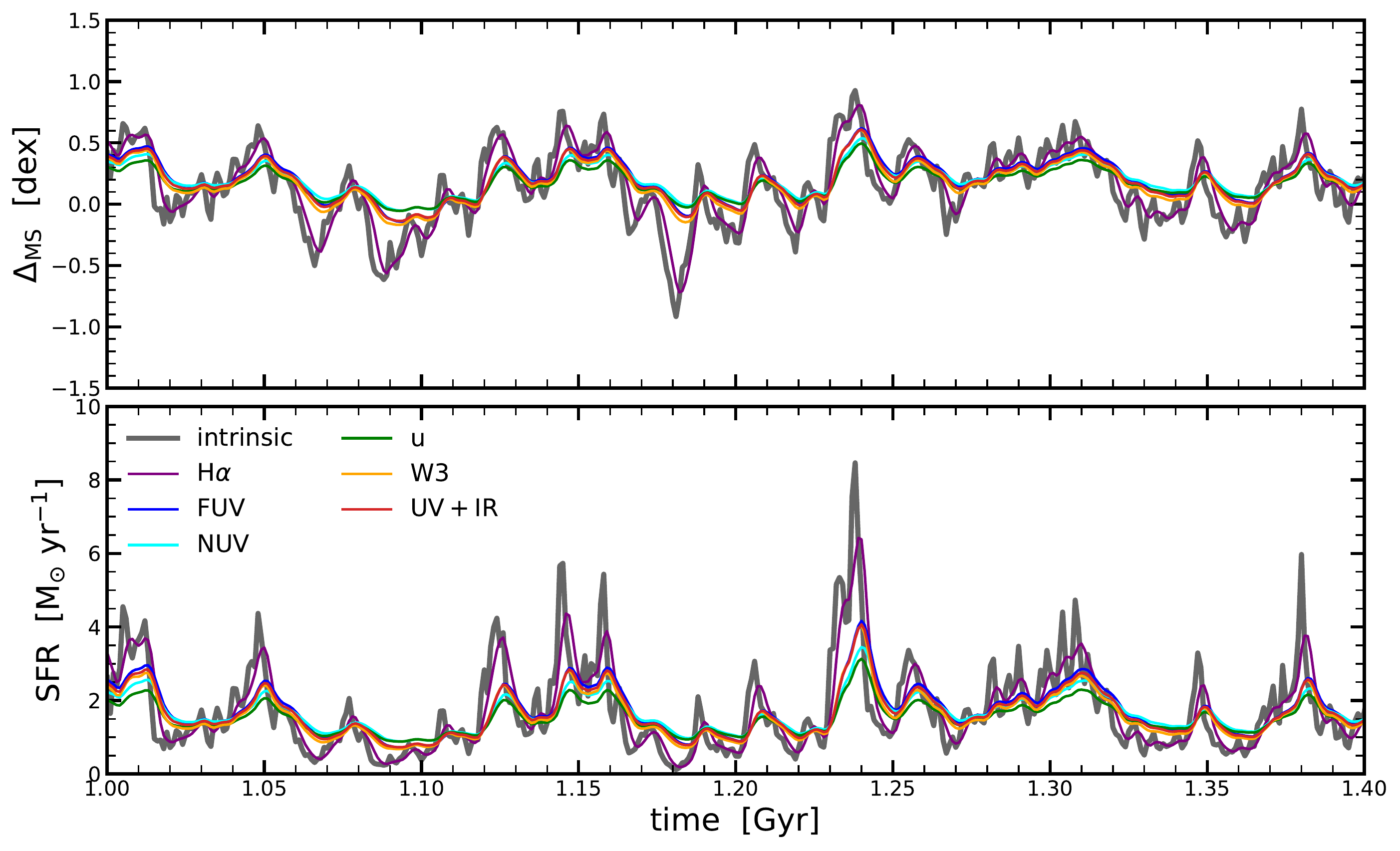}
    \caption{Example of the SFH of a galaxy as seen by different SFR indicators. The upper panel shows the distance from the MS as a function of time, while the bottom panel shows the absolute SFR, assuming the MS has a normalization of $1~M_{\odot}~\mathrm{yr}^{-1}$ at all times and stellar masses. We show a small zoomed-in section from one of the simulated SFH, starting at 1 Gyr. The intrinsic SFH is shown in black, while the purple, blue, cyan, green, orange and red lines show the SFH as seen by the H$\alpha$, FUV, NUV, u, W3, and UV+IR SFR indicators, respectively. Since H$\alpha$ probes short timescales (see Figure~\ref{fig:Levo}), it tracks the intrinsic SFH well. On the other hand, if the SFR indicator probes longer timescales, the overall SFH is much smoother and there is an overall bias towards higher SFRs, leading to an offset of the inferred MS ridgeline.} 
    \label{fig:SFH_indicator_example}
\end{figure*}

\begin{figure*}
    \centering
    \includegraphics[width=\textwidth]{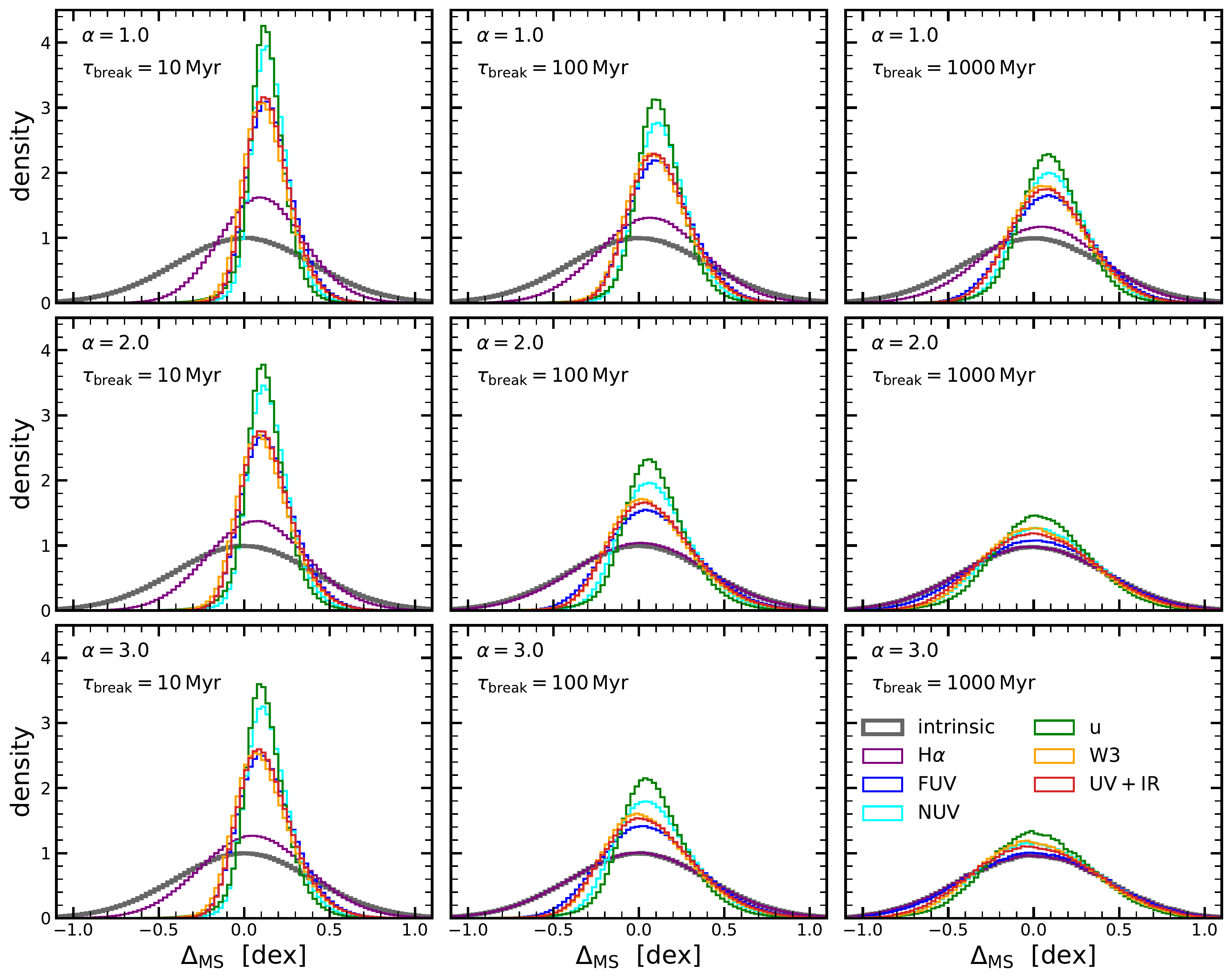}
    \caption{Distribution about the MS ridgeline as seen by different SFR indicators. This figure is similar to Figure~\ref{fig:DMS_distribution}, but this time for realistic SFR indicators. Each panel shows the resulting distributions for a given PSD, with increasing $\alpha$ and $\tau_{\rm break}$ from top left to bottom right. The thick black histogram indicates the intrinsic distribution, which has by construction a mean of 0.0 and a scatter of $0.4~\mathrm{dex}$ for all $\alpha$ and $\tau_{\rm break}$. The purple, blue, cyan, green, orange and red histograms show the distribution of galaxies as seen by the H$\alpha$, FUV, NUV, u, W3, and UV+IR SFR indicator, respectively. The larger the timescale probed by the indicator, the larger the MS offset and the smaller the MS scatter. The effects are the largest for small $\alpha$ and $\tau_{\rm break}$ since the PSD has a lot of power on short timescales relative to the long timescales.} 
    \label{fig:DMS_distribution_indicator}
\end{figure*}

\begin{figure*}
    \centering
    \includegraphics[width=\textwidth]{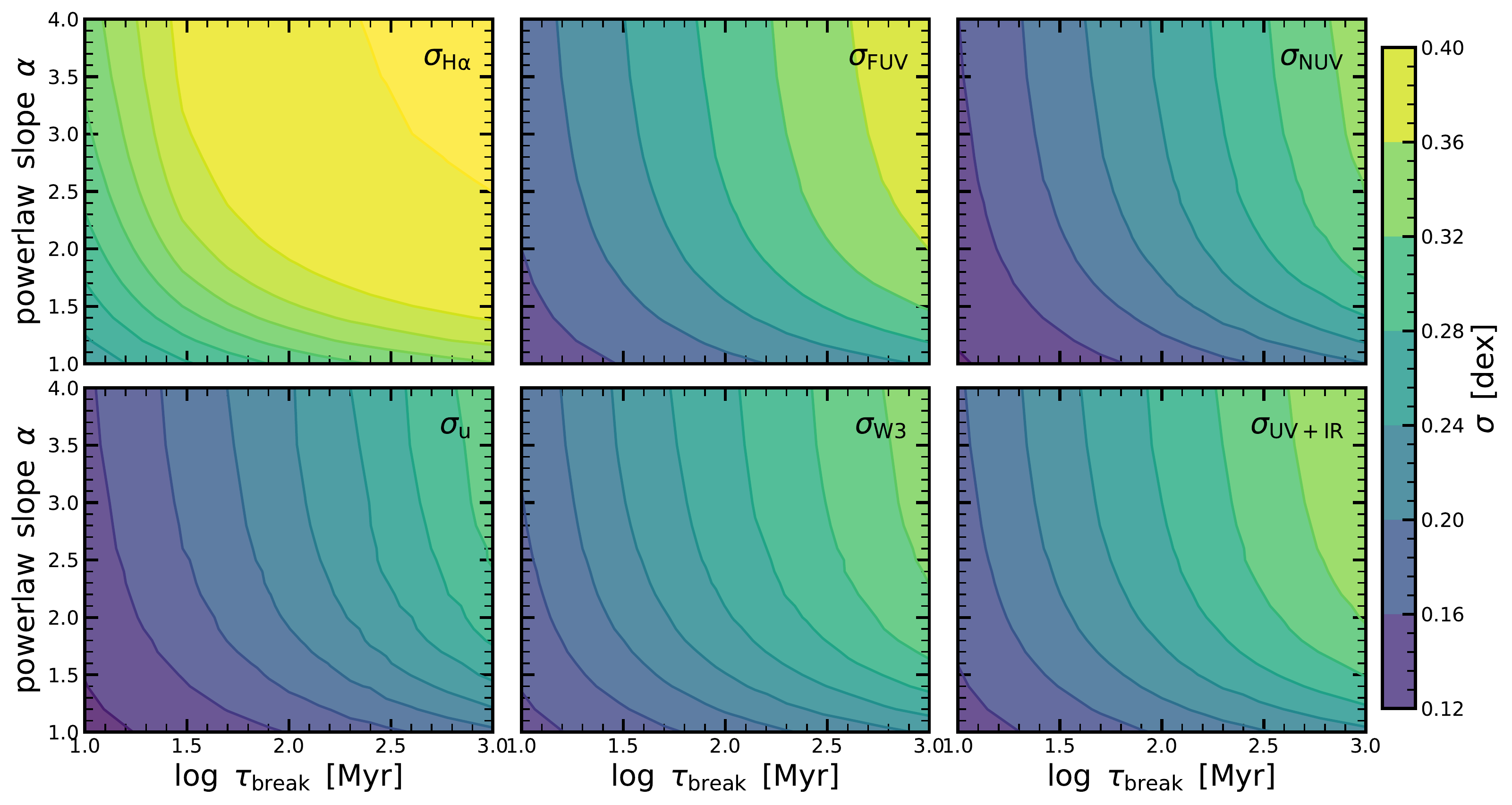}
    \caption{The scatter of the MS as seen by different indicators. Each panel shows the MS scatter as seen by a certain indicator as a function of the PSD parameters $\tau_{\rm break}$ and $\alpha$. The panels from the top left to the bottom right show the results for the H$\alpha$, FUV, NUV, u, W3, and UV+IR SFR indicator. In general, the burstier the SFR (see Figure~\ref{fig:Burstiness}), the smaller the measured scatter with the indicator.} 
    \label{fig:scatter_indicators}
\end{figure*}

\begin{figure*}
    \centering
    \includegraphics[width=\textwidth]{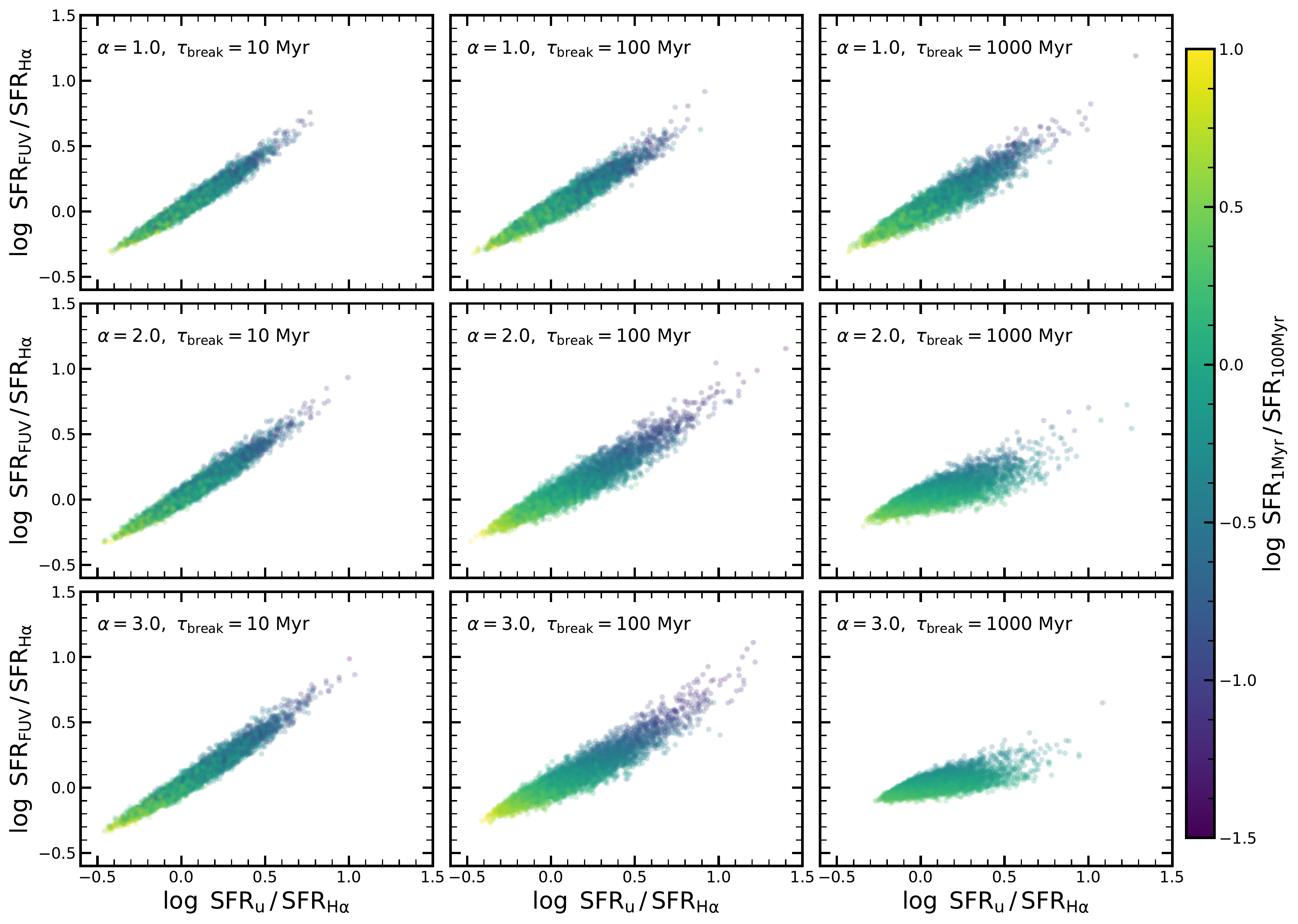}
    \caption{The distribution of galaxies in the plane of $\mathrm{SFR}_{\rm FUV}/\mathrm{SFR}_{\rm H\alpha}$ versus $\mathrm{SFR}_{\rm u}/\mathrm{SFR}_{\rm H\alpha}$. Each panel shows the resulting distributions for a given PSD, with increasing $\alpha$ and $\tau_{\rm break}$ from top left to bottom right. The color-coding corresponds to the ratio of the SFR measured over 1 Myr and the SFR measured over 100 Myr. The shapes of these distributions differ - highlighting the possibility to constrain the PSD with such observations. } 
    \label{fig:SFR_ratio}
\end{figure*}

We highlighted in the previous section that the MS scatter for the SFR average over different timescales can give a tight constraint on the PSD of the underlying stochastic process. In this section, we present a detailed investigation quantifying how does the normalisation and width of the MS change when probed with different SFR indicators, which themselves trace different timescales. In Section~\ref{sec:Observational}, we will use this information to show example of constraining the PSD from observations. \par

\subsection{SFR tracers}

There are multiple methods for determining SFRs from observational tracers such as from the UV, IR and emission lines. In the end, all the methods -- irrespective of whether the observable is directly starlight, dust-reradiated starlight or nebular line emission -- quantitatively link the observable to the intrinsic emission from hot young stars. Therefore, a crucial part of each method deals with how to correct for dust attenuation. Similarly, the IMF, in particular the high-mass end slope, and the star-formation history play a role in the exact luminosity-SFR calibration. \par

In the scope of this work, we are primarily interested in the distribution of SFRs probed by different tracers and hence timescales. We assume that measurements of dust-corrected SFRs as provided in the literature are accurately dust corrected, i.e., we do not discuss any dust effects. Similarly, we assume a \citet{chabrier03} IMF and solar metallicity throughout. Variations in metallicity, dust attenuation (in particular differential dust effect between young and old stars) and IMF could be, in principle, important. We will address this in an upcoming publication, where we will study the effect of variability of SFHs, IMF and dust attenuation in the observational plane of luminosities. In general, these effects would increase the scatter, implying that our results are basically an upper limit on the variability. In Section~\ref{sec:Observational}, we focus on a galaxy sample in a small mass range ($M_{\star}=10^{10}-10^{10.5}~M_{\odot}$) in order to minimize variations in IMF, metallicity and dust content. \par

We are considering the following standard SFR tracers \citep[see][]{kennicutt98,kennicutt12,davies16}:

\begin{itemize}
\item H$\alpha$: H$\alpha$ photons arise from the gas ionized by young ($<20$ Myr) stars. Thus, H$\alpha$ provides a measure of the current SFR in galaxies. 
\item FUV and NUV: UV continuum emission arises from hot, massive O and B stars with $M_{\star}>3~\mathrm{M}_{\odot}$, and hence is a good tracer of more recent star formation, though on longer timescales than H$\alpha$.
\item $u$-band: The rest-frame $u$-band ($\sim3500$ \AA) emission arises from the photospheres of young, massive stars, tracing the star formation on $\la100$ Myr timescales. This emission is less strongly affected by dust obscuration than the UV, but the possible large contribution from older stellar populations makes it more difficult to interpret.
\item \textit{WISE} W3: Probing infrared (IR) flux from star-forming galaxies in the $3-100~\mu\mathrm{m}$ range gives a reliable estimate for the ongoing star-formation: the amount of flux emitted in the IR is directly related to the UV emission from newly formed stars \citep[e.g.]{calzetti07}. We consider the \textit{Wide-field Infrared Survey Explorer} (\textit{WISE}) passband W3 at rest-frame $12~\mu\mathrm{m}$ to estimate the SFR \citep{cluver14}.
\item UV+IR: As discussed above, UV emission arises directly from star-forming regions, while some fraction of this emission is absorbed and reprocessed by dust, being re-emitted in the IR. We therefore sum both UV and total IR luminosities to obtain the total SFR, based on the bolometric luminosity of OB stars.
\end{itemize}
\par
In order to now predict what the SFH, as seen by a certain indicator looks like, we need to convolve the intrinsic SFH with the response function of that indicator.  Response functions for the aforementioned indicators are shown and discussed in Appendix~\ref{app:response_functions}. Briefly, we predict the luminosity evolution for a Simple Stellar Population (SSP) using the Flexible Stellar Population Synthesis code \citep[\texttt{FSPS}\footnote{\url{https://github.com/cconroy20/fsps}};][]{conroy09a}. \texttt{FSPS} has been extensively calibrated against a suite of observational data \citep[for details see][]{conroy10}. Throughout this work, we adopt the \texttt{MILES} stellar library and the \texttt{MIST} isochrones. Since the W3 and UV+IR SFR tracers are based on emission from a large wavelength range, these indicators depends on the dust attenuation prescription. We assume a uniform screen with \citet{cardelli89} attenuation curve and an opacity of $\tau_{\rm}=0.5$ at $5500~$\AA, motivated by findings by \citet{salim18_curves} in the local universe. Changing this prescription does not affect our findings here significantly. As expected, and shown in Appendix~\ref{app:response_functions}, the H$\alpha$ SFR indicator shows a rapid response function, followed by the FUV SFR indicator and the other indicators.\par

Figure~\ref{fig:SFH_indicator_example} shows the effect on the SFH by convolving the SFH with the response functions for the different indicators. In particular, the upper panel shows the distance from the MS, while the lower panel shows the absolute SFR under the assumption that the MS has a SFR of $1~M_{\odot}~\mathrm{yr}^{-1}$ at all times and masses. This SFH has PSD described with a high frequency power-law slope of $\alpha=2$ and a decorrelation timescale of $\tau_{\rm break}=100~\mathrm{Myr}$. \par

The SFR inferred from H$\alpha$ traces the intrinsic SFH well, only missing fluctuations on the shortest timescale ($\sim4$ Myr). All other tracers, which probe longer timescales, show a much larger differences compared to the intrinsic SFH. We find that the SFHs are not only smoother, but because of the long-timescale tail in the response function, the SFRs are biased high. There is an overall mean offset of the MS. Additionally, we find that all the variability on short timescales ($<20$ Myr) are washed out, giving rise to a narrower MS width. This is overall consistent with the simple analysis presented in the previous section.  \par

\subsection{Shape of the MS as seen by different SFR indicators: normalization and scatter}

We go now one step further and look how the effect on the MS depends on the PSD itself. Figure~\ref{fig:DMS_distribution_indicator} is a remake of Figure~\ref{fig:DMS_distribution}, showing the distribution of the distances from the MS ridgeline for different SFR indicators. We see the same qualitative behavior that we discussed extensively in Figure~\ref{fig:DMS_distribution}, when we discussed the simplified case in which we simply averaged SFRs over a certain timescale. Measurements in H$\alpha$, which probe star-formation on the shortest timescale, follows the intrinsic distribution closely for all but the most ``bursty'' set of parameters. On the other hand, all other indicators show some level of deviation, and produce a narrower distribution that is offset from the intrinsic value of the MS. Indicators that act on longer time-scales tend do produce narrower distributions, which is in very good aggrement with observations (e.g., Figure 7 from \citealt{davies19}). The differences in the shapes, even if not so pronounced as in the simple case shown in Figure~\ref{fig:DMS_distribution}, can be again used to infer the intrinsic properties of the underlying stochastic process.\par

Although both the normalisation as well as the width of the MS as seen by different indicators can in principle be used to derive the PSD, we are focusing now on the width of the MS. We do this because the overall normalisation depends on the exact conversion used to derive SFRs from fluxes, which are uncertain and depend on several assumption like the IMF, metallicity, as well as the assumed SFH.  On the other hand, the width of the MS is a relative measure that depends only on the rank-ordering in the galaxy sample and as such it is largely independent on the exact flux-SFR conversion. \par

Figure~\ref{fig:scatter_indicators} shows how the scatter of the MS changes when traced by different SFR indicators. This figure assumes that the intrinsic width of the MS is 0.4 dex (probed on a 1 Myr timescale). Specifically, each panel shows the MS scatter as a function of the PSD parameters $\alpha$ and $\tau_{\rm break}$. We see a similar pattern that we have also seen when we have discussed ``burstiness" (Figure \ref{fig:Burstiness}). Consistent with the previous figure, measurements in H$\alpha$ follow the intrinsic width closely for large parts of the parameter space. Measurements in other indicators that depend more heavily on the previous SFH provide a measurement which is less readily interpretable. In general, as discussed in Section \ref{sec:Description}, processes with low $\alpha$ and $\tau_{\rm break}$ that oscillate wildly, lead to tighter observed MS, courtesy of the central limit theorem. Although measured width in different indicators are relatively similar, the differences between them enable us to constraint parameters of the underlying process in Section \ref{sec:Observational}. \par

\subsection{Ratio of SFRs with different indicators}

We now move forward to a MS-independent discussion and directly compare different SFR tracers with each other in Figure~\ref{fig:SFR_ratio}. Specifically, we plot the ratio of FUV and H$\alpha$ SFRs versus the ratio of $u$-band and H$\alpha$ SFRs. The color-coding corresponds to the ratio of the SFR measured over 1 Myr and the SFR measured over 100 Myr. There are three trends visible from this figure. Firstly, the tightness of the relation increases with increasing burstiness. This arises because all SFR indicators probe the same SFR if the SFR varies on shorter timescales than the timescales of the SFR indicators. Secondly, the slope decreases with decreasing burstiness. This can be explained by the fact that the H$\alpha$ and FUV SFR indicators are able to probe short-term variability better than the $u$-band SFR indicator, which, in the case of slowly varying SFHs, leads to a FUV-to-H$\alpha$ ratio close to 1 while the $u$-band-to-H$\alpha$ ratio shows significantly more scatter. Thirdly, the color-coding shows that if the SFH quickly increases, H$\alpha$ is able to probe this, leading to ratios smaller than 1, while if the SFH quickly decreases, the ratios are all larger than 1. Overall, this figure highlights that -- in the framework presented in this work -- there are quantifiable consequences which arise when simply investigating the relation between the SFR measured with different indicators.\par

\section{Observational constraints} \label{sec:Observational}

In this section, we wish to apply our insights about different stochastic processes and their influence on observed MS properties. The main goal is to use the MS width measurements to determine parameters of the stochastic process, i.e., slope $\alpha$ and decorrelation timescale $\tau_{\rm break}$ of the PSD, that govern the evolution of star-forming galaxies about the MS ridgeline. The analysis in this section should be regarded only as a proof of concept as we work with the simplest possible assumptions, in order to highlight the power of our approach. We show that the result for the parameters that we derive are consistent when using different indicators, which is highly encouraging.  \par

\subsection{Main result}

The ideal dataset should contain statistically significant sample of galaxies and be covered in a several observational bands, in order to check consistency between different SFR indicators. Additionally, the analysis is simplified in the case of a galaxy sample at low redshift. This is because of (a) the slow evolution of the MS as a function of time, which simplifies interpretation of measured SFRs as an offset from the virtually non-changing MS; and (b) the slow growth of mass, meaning that measurement of total stellar mass and star-formation are independent quantities. We discuss these issues more in Section \ref{sec:Discussion}. Additionally, low redshift sample also avoids potential problems and systematics that might arise due to K-correction.\par

We find the measurement of the MS width presented in \citet{davies19} to be well suited for this analysis. \citet{davies19} used data from Galaxy and Mass Assembly Survey \citep[GAMA;][]{driver11, liske15, driver16, baldry18} and conducted an analysis which was explicitly measuring width of the MS with different SFR indicators. The sample consists of 9005 galaxies and is restricted to $z<0.1$, satisfying our requirement to work with a large sample at low redshift. It includes only galaxies which are not classified as being a member of a group or a pair in the GAMA group catalogue, as described in \cite{robotham11}. This means that our final result will be more suited to describing star-formation fluctuations due to ``secular'' evolution instead of mergers.\par
	
We consider the following measurements: (a) extinction-corrected H$\alpha$ SFRs, measured with the process outlined in \cite{gunawardhana11, hopkins13, gunawardhana15} and using line measurements from \cite{gordon17}, (b) UV+IR SFRs, derived from the galaxy spectra as described in \cite{brown14}, and (c) extinction-corrected $u$-band SFRs from rest-frame u-band luminosity \cite{davies16}. SFRs have been derived using Chabrier IMF. We refer the reader to \cite{davies19} for more details.\par

We consider MS width measurements for SFR selected galaxies between $10^{10.0}$ and $10^{10.5}~M_{\odot}$ as we expect that the measurement errors are the smallest for this subset of galaxies. Furthermore, more massive galaxies are transitioning from the star-forming to the quiescent population \citep[quenching, e.g.,][]{peng10_Cont}, which could lead to an artificially large scatter of the MS, which itself would depends heavily on the selection of star-forming galaxies. We estimate, using Equation (27) from \citet{peng10_Cont} with reasonable quenching time of 1 Gyr, that only around 2$\%$ of star-forming galaxies at this mass are being ``quenched'' at redshift 0.  Additionally, considering the rather narrow mass range makes our result less affected by possible variations of dust and IMF within the sample. In order to then derive the observational constraints, we assume a relative error on the ratios of $7\%$ to include possible observational uncertainties in the error budget \citep{davies19}.\par

\begin{figure}
    \centering
    \includegraphics[width=\linewidth]{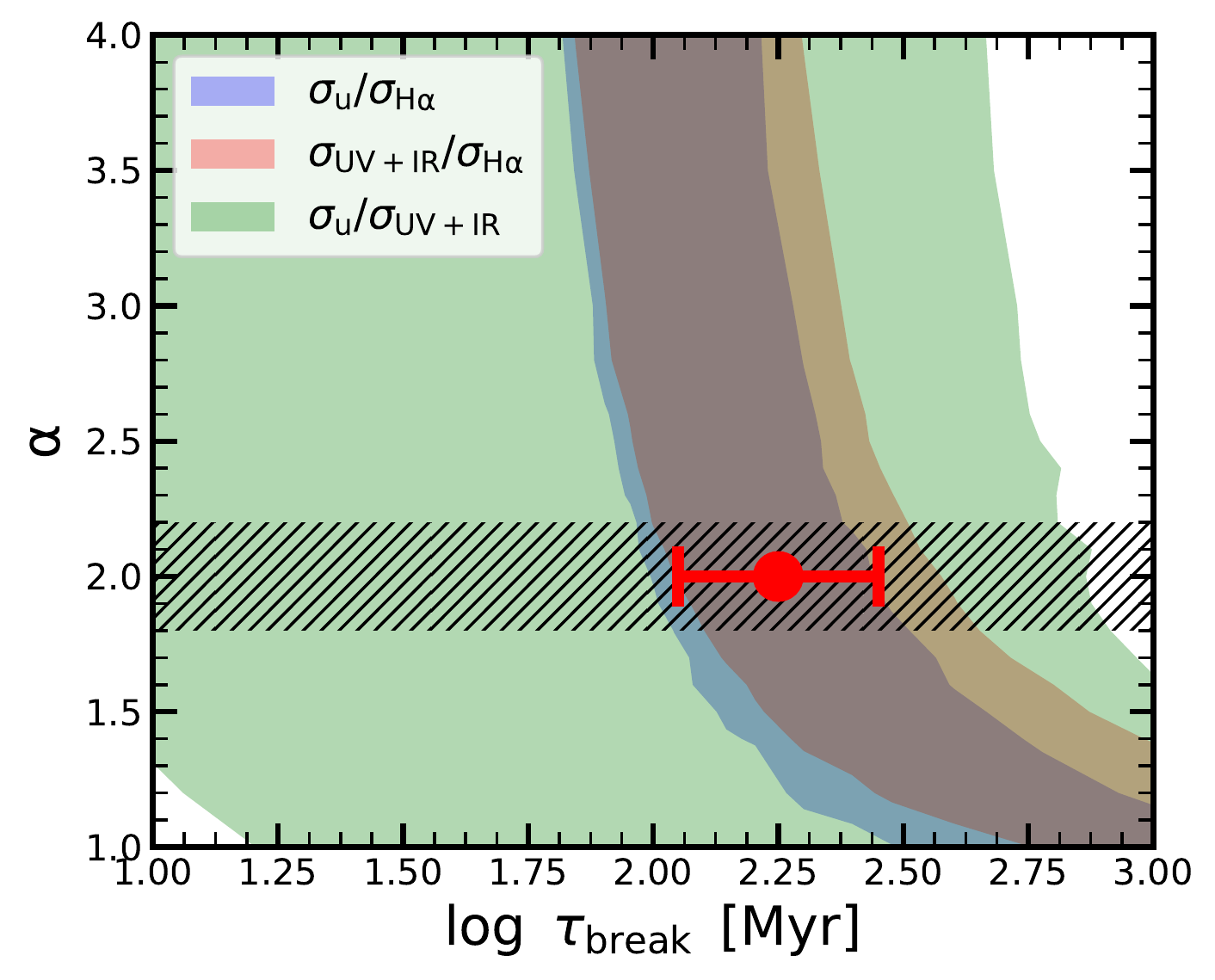}
    \caption{Observational constraints on the PSD. We plot current observational constraints from the width of the MS in the plane of $\alpha$ and $\tau_{\rm break}$. Specifically, the blue, red and green shaded regions show the constraints obtained from the ratio of the width of the MS as measured using the $u$-band and H$\alpha$, UV+IR and H$\alpha$, and $u$-band and UV+IR, respectively. The overlapping region gives a tighter constraint in $\tau_{\rm break}$ than $\alpha$.  Assuming $\alpha\approx2$, we obtain $\tau_{\rm break}=178^{+104}_{-66}~\mathrm{Myr}$, which is indicated by the horizontal red errorbar.} 
    \label{fig:observational}
\end{figure}

The final result is shown in Figure \ref{fig:observational}. As elaborated in Section \ref{sec:Finding}, we consider ratios of the MS widths measured with different indicators to reduce the influence of the unknown intrinsic width of the MS on our measurement. We see the similar structure as in Figure \ref{fig:not_degenerate}, with the general degeneracy between $\alpha$ and $\tau_{\rm break}$. We wish to highlight the consistency between the parameter space covered by all three combination of the indicators used, albeit the ratio $\sigma_{\rm u,MS}/\sigma_{\rm UV+IR,MS}$ is quite uninformative as it covers large part of the considered parameters space. Nevertheless, we find the consistency between all of the indicators very encouraging, especially given the simplicity of our assumptions and minimal consideration of the dust and IMF effects.\par

Unfortunately, with this data and simplicity of our analysis is not possible to resolve the degeneracy between  $\alpha$ and $\tau_{\rm break}$, as it has been possible in the simplified case considered in Section \ref{sec:Comparisontonumerically}. This is due to combination of two factors. Firstly, this is due to larger uncertainties in the observational measurement of the $\sigma_{\rm MS}$, driven by the smaller number of galaxies in the sample and by the observational uncertainties. Secondly, fact that realistic indicators are not a simple step functions but have non-trivial time dependence that mixes many timescales reduces our ability to differentiate between the influence of $\alpha$ and $\tau_{\rm break}$ in observations (see Figure \ref{fig:scatter_indicators} and Appendix \ref{app:response_functions}). \par

In order to quantify our final result in one number, we report $\tau_{\rm break}$ after setting the high-frequency slope at $\alpha=2$, i.e., for the case of damped random walk. We emphasize however that the data itself does not prefer $\alpha=2$ over any other particular value for $\alpha$ and our choice for reporting $\tau_{\rm break}$ for damped random walk is driven by theoretical and empirical reasons outside of the scope of this paper. The first reason for considering the damped random walk is its theoretical and physical simplicity. Secondly, the slope of $\alpha=2$ is seen when studying time-variability properties of galaxies in simulations (Iyer et al. in prep.). In the case of $\alpha=2$, we find that the region of the $\tau_{\rm break}$ parameter space that is within 1-$\sigma$ contours for all combinations of indicators is given with $\tau_{\rm break}=178^{+104}_{-66}$ Myr. We intentionally do not conduct further statistical consideration of the allowed region, given that we believe that systematic uncertainties, due to variation of dust, IMF and observational uncertainties dominate our error budget and would make any such analysis highly uncertain. We do note that measurement of the width of MS by other indicators available in \cite{davies19} are in broad agreement with the estimated value that is reported above. \par
 
\subsection{Discussion} \label{sec:Discussion}

\subsubsection{Mass dependence} 

The result of $\tau_{\rm break}=178^{+104}_{-66}~\mathrm{Myr}$ that we have derived is only valid for the sample that we considered, consisting of isolated galaxies with $\log(M_{\star}/M_{\odot})=10.0-10.5$. As such, the derived timescale is not necessarily the same for the whole range of masses and environments in the Universe. We do note that there is a mass dependence in the measurements of the width of the MS in \cite{davies19}. We do not conduct the mass dependence analysis in this work given the possible mass dependence in the dust and IMF properties. We do however note that the measured widths of the MS suggest similar or somewhat shorter timescale for $\tau_{\rm break}$ for the smaller masses within the dataset ($8 <\log(M/M_{\odot}) < 10$), with some dependence on the exact set of indicators used.\par

\subsubsection{Caveats} \label{sec:Caveats}

Of course, our analysis comes with several caveats. One complication that we consider is the intrinsic correlation between the stellar mass and measured SFR. As we have elaborated, the measured SFR can be considered as an imperfect averaging over the intrinsic SFR, where the length of this averaging is dependent on the indicator used for the measurement. On the other hand, the stellar mass of a galaxy is also obviously related with the previous SFR, given by the integral over the complete SFH of the galaxy. Hence, the measurement of the MS width is only meaningful if the total stellar mass added over the period that a given indicator is measuring is small compared to the total stellar mass of a galaxy. Given that the mass doubling time for star-forming galaxies today is very long, i.e., comparable to Hubble time, the mass added over the time effectively probed by the different indicators is negligible (see also Appendix \ref{sec:Timescales}). This effect, however, needs to be accounted for at higher redshifts (e.g., $z>1$) where the timescale of the measurement of star-formation in slow-response indicators starts being a non-negligible fraction of the lifetime of the Universe \citep[e.g.,][]{tacchella18}.\par

Similarly, when conducting our analysis of observational results we have assumed that the overall change of the mean SFR of the MS is negligible. Consider the timescale of 400 Myr, which is more than twice as long as the $\tau_{\rm break}$ derived above. We show in Appendix \ref{app:response_functions} that by this timescale all of the considered indicators have outputted more than 60\% of their complete luminosity, i.e., are very weakly sensitive to the star-formation on longer timescales. If we assume relatively steep evolution of the sSFR of the MS, such as $(1+z)^{3}$ \citep[e.g.,][and references therein]{lilly13_bathtube}, we find that the ratio between $\mathrm{sSFR}_{\rm MS} (400 \mbox{ Myr})/\mathrm{sSFR}_{\rm MS}(\mbox{today}) \approx 1.1$ . As such this change will have minimal effect on the inferred result in this study, but it needs to be considered at higher redshifts. \par

Finally, observational uncertainties introduce scatter and, hence, increase the MS width obtained by \citet{davies19}. These uncertainties arise, beside other things, from  the aperture correction, the dust attenuation correction, and the assumption that all galaxies have the same IMF and metallicity. It is beyond the scope of this paper to assess the impact of those uncertainties in detail, but these all contribute to relatively large error on measured $\sigma_{\rm ind, MS}$ that we have used. \par

 \section{Summary and conclusions} \label{sec:Summary}

The knowledge of how the SFR changes as a function of time for individual galaxies is a critical ingredient needed for understanding the evolution of galaxies. SFHs encode information about the variability on short and long timescales that arise from different physical processes, such as gas accretion, mergers, and feedback from stars and black holes. The best method to model, and mathematical describe, SFHs is is still being debated. \par

In this paper, we present a framework for modelling SFHs of galaxies as a stochastic process. We focus on star-forming galaxies and assume that a stochastic process describes how these galaxies move about the MS ridgeline. We also assume that at any moment the intrinsic distribution of SFRs around the MS ridgeline is log-normal. We define this stochastic process through a power spectrum density. The functional form of the PSD is given by a broken power-law, with high-frequency slope $\alpha$ that flattens out at a frequency corresponding to $\tau_{\rm break}$. The physical meaning of the PSD is that SFHs are correlated on short timescales, where the strength of this correlation is described by the slope $\alpha$, and they decorrelate to resemble white noise at a timescale that is proportional to $\tau_{\rm break}$. Specifically, when the slope of the PSD is shallow and/or $\tau_{\rm break}$ is small, the SFH violently oscillates around the mean and is very quickly decorrelated. On the other hand, steeper slopes and longer $\tau_{\rm break}$ lead to slower oscillations and a more correlated behavior. \par

We demonstrate that, even though we can not observationaly follow the time-evolution of individual galaxies, we can deduce the properties of the stochastic process. This can be done by measuring properties of the MS, such as the MS normalisation and the MS scatter, with different observational indicators that are sensitive to different timescales. There exists a degree of degeneracy between the influence of the parameters $\alpha$ and $\tau_{\rm break}$, that can be roughly described by the ``burstiness'', i.e., by the fact that galaxies described with smaller $\alpha$ and shorter $\tau_{\rm break}$ tend to more often have SFRs which are outside of the 1-$\sigma$ width of the MS. We show that this degeneracy can be broken, in principle, by measuring the properties of the MS in several observational indicators. \par 

We then consider realistic observational SFR indicators, and show results for the measured properties of the MS in H$\alpha$, FUV, NUV, $u$-, W3 and UV+IR bands. We use those results to, as a proof of concept, deduce parameters of the underlying stochastic process from the data from the GAMA survey, presented in \cite{davies19}. For a sample of isolated galaxies with $\log(M_{\star}/M_{\odot})=10.0-10.5$ and using MS widths measured from H$\alpha$, $u$- and UV+IR bands, we find the area of the parameter space that is consistent with this data. With an assumption that the high-frequency slope is given with $\alpha=2$, we obtain $\tau_{\rm break}=178^{+104}_{-66}$ Myr. The motivation for $\alpha=2$ stems from numerical simulations as well as the theoretical attractiveness of the damped random walk process. The quoted errobars are deduced from the observational 1-$\sigma$ errors of all the available indicators. \par

Our result of $\tau_{\rm break}\approx200$ Myr indicates that the SFHs of galaxies decorrelate, i.e., lose memory of their previous SFH, on roughly this timescale. This is shorter than the dynamical time of the dark matter halo, which is roughly 10\% of the Hubble time (i.e., a bit more than a Gyr at $z\sim0$). Therefore, we conclude that baryonic effects within the halos, which act on dynamical timescales of galaxies, play an important role in shaping the SFHs of galaxies. Examples of such effects could include  feedback and reincorporation of galactic winds. \par

In the future, we will apply this framework to a wide range of numerical and semi-analytical models, where we can measure the PSD directly of the models' SFHs. Preliminary results show that the obtained PSDs are well described by a broken power-law with nearly all theoretical models producing $\alpha\approx 2$, while having different values for $\tau_{\rm break}$. We show that the PSD is set by physical ingredients in the models and hence that the PSD contains a wealth of information that has not yet been researched. On the observational side, we will address many issues that complicate precise determination of the PSD in observations, including possible variations in the IMF and dust attenuation. There we also wish to use these insights to analyse properties of galaxies at higher redshifts, where it is especially important to consider these complications.\par

\section*{Acknowledgements}
We thank Luke Davis for sharing his measurements from the GAMA survey with us. Furthermore, we are thankful to Charlie Conroy, Luke Davis, Daniel Eisenstein, Kartheik Iyer, Ben Johnson, Daniel Kelson, Joel Leja, and Michael Strauss for very useful comments. We thank Sophie Reed and Hassan Siddiqui for proofreading the manuscript.  This research made use of NASA's Astrophysics Data System (ADS), the arXiv.org preprint server, the Python plotting library \texttt{matplotlib} \citep{hunter07}, \texttt{astropy}, a community-developed core Python package for Astronomy \citep{astropycollaboration13}, and the python binding of \texttt{FSPS} \citep{foreman_mackey14}. S.T. is supported by the Smithsonian Astrophysical Observatory through the CfA Fellowship.

\bibliographystyle{mnras}
\bibliography{library}

\appendix

\section{Fractional Brownian motion and Hurst parameter} \label{sec:Hurst}

\begin{figure}
	\centering
    \includegraphics[width=0.99\linewidth]{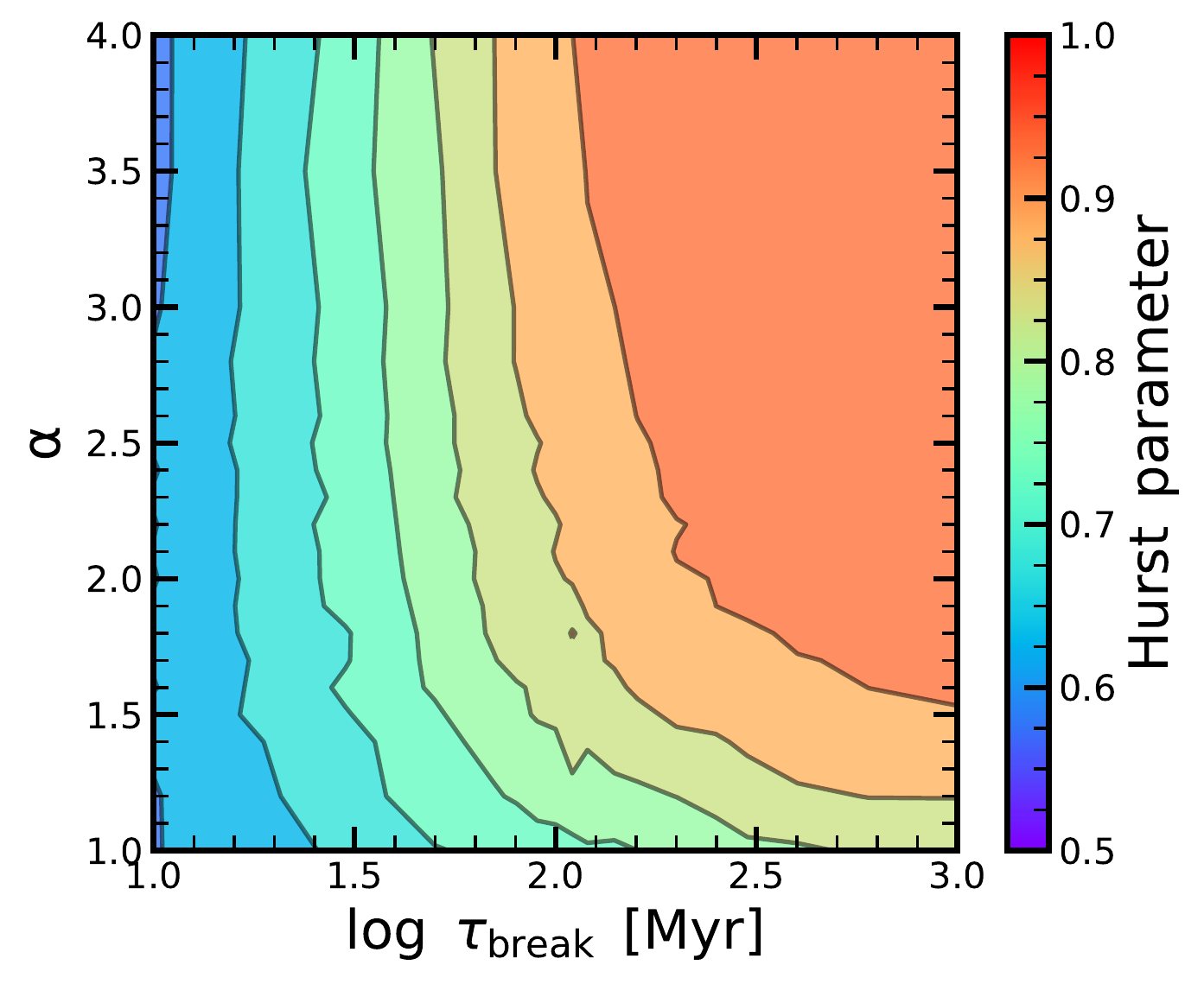}
    \caption{Hurst parameter, $H$, in the plane of the break timescale $\tau_{\rm break}$ and power-law slope $\alpha$ of the PSD. $H$ increases towards $\sim1$ with increasing $\tau_{\rm break}$ and $\alpha$, consistent with the SFHs shown in Figures~\ref{fig:Example_SFH} and \ref{fig:Example_PSD}. At $\alpha>2$ and $\log\tau_{\rm break}/\mbox{Myr} >1.5$, the Hurst parameter converges to $\sim1$ and is not able to quantify any differences in the PSD. }
    \label{fig:Hurst}
\end{figure}

In this appendix, we shortly discuss the connection between the  description of the stochastic processes that we have used in this work and the fractional Brownian motion described with the Hurst parameter, as used in \cite{kelson14}\footnote{Here we discuss ``type 2'' fractional Brownian motion, which is more suited for descriptions of physical process as in that description the future events are only dependent on the previous events. ``Type 1'' processes are dependent on both future and past events, violating causality.}. Fractional Brownian motion is process described with the ACF \citep{marinucci99}:

\begin{equation}
ACF(t)=\frac{1}{2} \left[-(t)^{2H} g\left(\frac{1}{t}\right)+(t+1)^{2H}+(t-1)^{2H} g\left(\frac{1}{t-1}\right)-t^{2H}\right]
\end{equation}
where 
\begin{equation}
g(t^{'})=1+2H \int^{t^{'}}_{0} \left( (1+s)^{H-1/2}-s^{H-1/2} \right)^{2} ds,
\end{equation}

\noindent
and $H$ is called the ``Hurst exponent'' or ``Hurst parameter'' and it is bound between 0 and 1. In the limit when $t\rightarrow \infty$ the relation simplifies to \citep{marinucci99}:
\begin{equation}
ACF(t)=\frac{1}{2} \left( (t+1)^{2H}-2(t)^{2H}+(t-1)^{2H}\right).
\end{equation}
\citet{tarnopolski16} shows that when $H>1/2$, which describes a physically interesting case where observations are more correlated at shorter timescales, the autocorrelation function behaves as $ACF (t) \propto t^{2-2H}$. The fact that the ACF decreases slower than $t^{-1}$ is often expressed as a statement that the process has a ``long-range'' dependence. One can compare this to ACFs that we present in right panel of Figure \ref{fig:Example_PSD} (see also Equation \eqref{eq:exampleACF}), which all reach $ACF (t) \sim 0$ at timescales correlated with $\tau_{\rm break}$. In other words, by modelling the PSD as broken power law, where the small-frequency slope is zero (mimicking white noise), we have explicitly allowed for the system to decorrelate.\par

Given that the processes that we use in this work have an explicit timescale built into them through the $\tau_{\rm break}$ parameter, we see that any connection that we want to derive between these two different approaches (fractional Brownian motion and broken-law PSD) will fundamentally depend on the timescale over which the connection is measured. In any case, it is crucial to note that any such a connection is essentially only an approximation, given the different definitions of both processes. In Figure \ref{fig:Hurst} we show the Hurst parameter which best approximates the SFHs generated with different $\alpha$ and $\tau_{\rm break}$. The SFHs used here have a length of 4.2 Gyr as described in Section \ref{sec:Generating}. We see similar dependence as in when we have discussed ``burstiness'' and widths of the MS measured in different indicators. As expected, a larger Hurst parameter is found for SFHs that are well correlated (larger $\tau_{\rm break}$ and $\alpha$) while less correlated, ``bursty'' SFHs (smaller $\tau_{\rm break}$ and $\alpha$) lead to lower Hurst parameters. \par

\section{Connection between star-formation measurements in different indicators} \label{sec:ConnectionAnal}

In this appendix we extend our analytical calculation of the measured width of MS, as seen after measurement in different indicators. We start by noticing that the star formation rate measured by a given indicator is connected to the previous star formation and is given by the intensity of that star-formation episode and a response of that particular indicator to the star formation, i.e.,
\begin{equation} \label{SFR_ind}
\mathrm{SFR}_{\rm ind}(t)=\mathrm{SFR}(t') f_{\rm ind}(t-t'),
\end{equation}
where $\mathrm{SFR}_{\rm ind}$ is the measured star-formation, $\mathrm{SFR}$ is the actual star-formation, and $f_{\rm ind}$ are response functions (the ones that we have used in this work are presented in Appendix \ref{app:response_functions}).
In the case of continuous star-formation the measured star-formation is given by
\begin{equation}
\mathrm{SFR}_{\rm ind}(t)=\int^{t}_{-\infty} \mathrm{SFR}(t')  f_{\rm  ind}(t-t') dt',
\end{equation}
where we have convolved the total star-formation history with an appropriate response function up to the time of the measurement.
When discretized the above integral is written as a sum
\begin{equation} \label{SFRind}
\mathrm{SFR}_{\rm ind}(t_{i})= \sum^{\infty}_{t_{j}=0} \mathrm{SFR}(t_{i}-t_{j}) w_{\rm  ind}(t_{j}),
\end{equation}
where $SFR_{\rm ind}(t_{i})$ is the measured star-formation in bin $t_{i}$ and $w_{\rm ind}(t_{j})$ is the discretized version of the response function:
\begin{equation}
w_{\rm  ind}(t_{j}) = \int^{t_{j}+\delta/2}_{t_{j}-\delta/2} f_{\rm ind}(t') dt',
\end{equation}
where $\delta$ is the duration of the discretization bin. In this work we use $\delta$ = 1 Myr (see Section \ref{sec:Description} for justification of this time-scale). \par

Now, consider the ensemble of galaxies moving around Main-Sequence in stochastic fashion with a log-normal distribution function. Lets denote with $\Delta$\footnote{As mentioned in the main body of the text, we use just $\Delta$ as shorthand for $\Delta_{\rm MS}$ in order to reduce the size and complexity of our expressions.} $\log$-distance from the median value of MS i.e., (see Equations \eqref{definitionDelta} and \eqref{definitionDeltaind}):
\begin{equation}
\Delta =\log \mathrm{SFR} - \log \mathrm{SFR}_{\rm MS}.
\end{equation}
\noindent
We rewrite the  Equation \eqref{SFRind} in terms of $\Delta$: 
\begin{equation} \label{masterSFRequation}
\Delta_{\rm ind}(t_i)= \log \sum^{\infty}_{t_j=0} 10^{\Delta (t_i-t_j)} w_{\rm ind}(t_j).
\end{equation}
 We note that that factors in the exponent $\Delta(i-j)$ are in general expected to be small as they are bound by the width of the MS. In that case we can use the MacLauren expansion:
\begin{equation} \label{eq:expansion}
10^{x}= 1+\ln(10)x+\frac{1}{2}\ln(10)^{2}x^{2}+\frac{1}{6}\ln(10)^{3}x^{3}+...
\end{equation}
Lets continue by taking first two orders of expansion above in Equation \eqref{eq:expansion}:
\begin{equation}\begin{split}
\Delta_{\rm ind}(t_{i})&=\log \sum^{\infty}_{j=0}  \left( 1  + \ln (10)\Delta(t_{i}-t_{j}) \right. \\& \left.+ \frac{1}{2}\ln(10)^{2}\Delta^{2}(t_{i}-t_{j})+....\right) w_{\rm ind} (t_{j})\\
&=\log  \left[  \left( \sum^{\infty}_{j=0}w_{\rm ind}(t_{j})\right)+\ln(10)\left(\sum^{\infty}_{t_{j}=0} \Delta(t_{i}-t_{j}) w_{\rm ind}(t_{j})\right) \right. \\& \left.+ \frac{1}{2}\ln(10)^{2} \left(\sum^{\infty}_{t_{j}=0} \Delta^{2}(t_{i}-t_{j})w_{\rm ind}(t_{j})\right)+....\right] \\
&=\log \left[1+  \ln(10) \sum^{\infty}_{t_{j}=0}  \Delta(t_{i}-t_{j}) w_{\rm ind}(t_{j}) \right.  \\& \left. +  \frac{1}{2}\ln(10)^{2} \left(\sum^{\infty}_{t_{j}=0} \Delta^{2}(t_{i}-t_{j})w_{\rm ind}(t_{j})\right)+....\right].
\end{split}\end{equation}
\noindent
where we have used the property that the response function has to sum to unity, i.e., $ \sum^{\infty}_{t_{j}=0}w_{\rm ind}(t_{j})=1$. We now expand the $\log$ function using
\begin{equation}
\log(1+ x)=\frac{x}{\ln(10)}-\frac{x^{2}}{2 \ln(10)}+\frac{x^{3}}{3\ln(10)} +...
\end{equation}
and we find the expression
\begin{equation} \begin{split}\label{eq:expansionTwo}
\Delta_{\rm ind}(t_{i})= & \sum^{\infty}_{t_{j}=0} \Delta(t_{i}-t_{j}) w_{\rm ind}(t_{j}) +  \frac{\ln(10)}{2} \left[ \left(\sum^{\infty}_{t_{j}=0} \Delta^{2}(t_{i}-t_{j})w_{\rm ind}(t_{j})\right)\right.\\& \left.-\left( \sum^{\infty}_{t_{j}=0} \Delta(t_{i}-t_{j}) w_{\rm ind}(t_{j}) \right)^{2} \right] +...
\end{split}\end{equation}
Let us now find variance of the measured MS after considering only the first term of the expansion above :
\begin{equation}\begin{split}\label{eq:expansionThree}
\sigma^{2}_{\rm ind,MS} (t_i)&=\mbox{VAR} \left( \sum^{\infty}_{t_{j}=0}\Delta(t_{i}-t_{j})w_{\rm  ind}(t_j)\right)  \\
&= \sum^{\infty}_{t_{j}=0} \sum^{\infty}_{t_{j^{'}}=0}  E\left( \Delta(t_{i}-t_{j})w_{\rm ind}(t_{j}) \Delta(t_{i}-t_{j^{'}})w_{\rm ind}(t_{j^{'}})\right)\\& - E\left( \Delta(t_{i}-t_{j})w_{\rm ind}(t_{j}) \right) E\left( \Delta(t_{i}-t_{j^{'}})w_{\rm ind}(t_{j^{'}}) \right) \\
& =  \sum^{\infty}_{t_{j}=0} \sum^{\infty}_{t_{j^{'}}=0} w_{\rm ind}(t_{j}) w_{\rm ind}(t_{j^{'}}) E\left( \Delta(t_{i}-t_{j}) \Delta(t_{i}-t_{j^{'}})\right) \\
& = \sum^{\infty}_{t_{j}=0} \sum^{\infty}_{t_{j^{'}}=0} w_{\rm ind} (t_{j}) w_{\rm ind}(t_{j^{'}}) \cdot COV \left(\Delta(t_{i}-t_{j})  ,\Delta(t_{i}-t_{j^{'}})\right) \\
& =  \sigma^{2}_{\rm MS} \sum^{\infty}_{t_{j}=0}\sum^{\infty}_{t_{j^{'}}=0} w_{\rm  ind}(t_{j}) w_{\rm ind}(t_{j^{'}})  ACF(t_{j}-t_{j^{'}}), 
\end{split}\end{equation}
where we have used the fact that the co-variance of the process is given by $COV(\Delta(t_{i}),\Delta(t_{j})$ = $\sigma^{2}_{\rm MS} ACF(t_{i}-t_{j})$ and where we have used symbol $E$ to denote expectation value. We have also used the definition of the covariance together with the fact that the expectation value of a stationary process equals zero: 
\begin{equation} \begin{split}
COV\left(\Delta(t_{i}-t_{j}) \Delta(t_{i})\right)  & =E\left(\Delta(t_{i}-t_{j}) \Delta(t_{i})\right)  -E\left(\Delta(t_{i}-t_{j})\right)  E\left(\Delta(t_{i})\right)  \\
& =E\left(\Delta(t_{i}-t_{j}) \Delta(t_{i})\right) .
\end{split}\end{equation}
Taking into the account next order terms in Equation \eqref{eq:expansionTwo} will lead to extra terms in the summation \eqref{eq:expansionThree}, with the smallest order being proportional to $\sigma^{4}_{\rm MS} $. This is because all $\sigma^{3}_{\rm MS} $ terms will inevitably be connected with terms $E(\Delta^{2}(t_{i}))\cdot E(\Delta(t_{j}))$ or $E(\Delta(t_{i}))\cdot E(\Delta^{2}(t_{j}))$ which are equal to zero, given the stationary of our problem. For the realistic widths of MS ($\sigma_{\rm MS} \sim $0.4$\pm 0.1$) these $\sigma^{4}_{\rm MS}$ terms are sub-dominant and therefore lead only to relatively small corrections (e.g., see the relatively small difference between the input and recovered parameters in Figures \ref{fig:distribution_single} and \ref{fig:not_degenerate}). \par

In the simplest case, discussed in Section \ref{sec:Finding}, in which we assume that a given SFR indicator is given by a step function lasting $N_{j}$ time units, i.e.,
\begin{equation*}
w_{\rm ind}(t_{j}) = \begin{cases}
1/N_{j} & 1 \leq t_j \leq N_{j} \\
0 &\text{otherwise.}
\end{cases}
\end{equation*}
the last expression in Equation \eqref{eq:expansionThree} can be written as 
\begin{equation}
\sigma^{2}_{\rm ind,MS} (t_{i})=\frac{\sigma^{2}_{\rm MS}}{N^{2}_{j}} \sum_{t_{j}=1}^{N_{j}}\sum_{t_{j^{'}}=1}^{N_{j}} ACF (t_{j}-t_{j^{'}}),
\end{equation}
and we can now recognize Equation \eqref{eq:VarEquation} used in the main body of the text.

\section{Luminosity evolution of a Simple Stellar Population (SSP)}
\label{app:response_functions}

\begin{figure*}
    \centering
    \includegraphics[width=\textwidth]{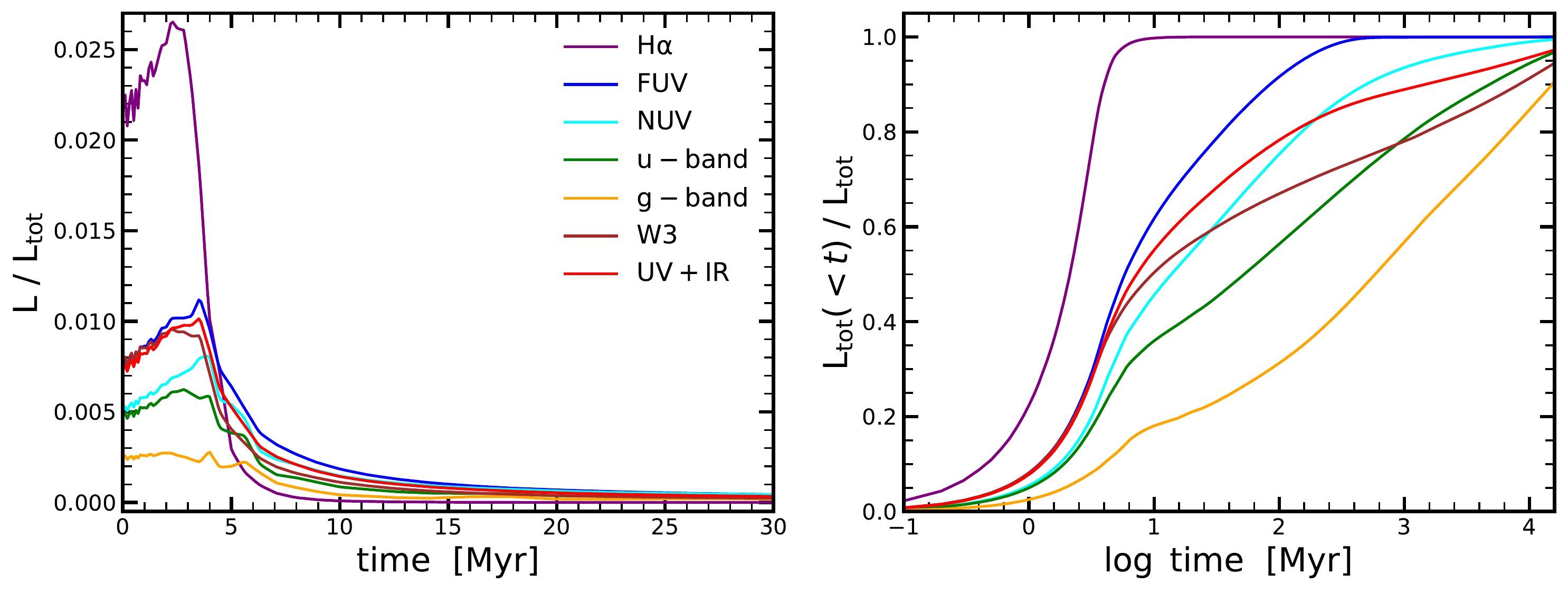}
    \caption{Luminosity evolution of different SFR tracers for a Simple Stellar Population (SSP). The left panel shows the luminosity as a function of time ($L(t)$), in bins of 0.1 Myr, normalized by total luminosity emitted over 25 Gyr ($L_{\rm tot},~ L_{\rm tot}\equiv \int^{25\mathrm{Gyr}}_{0}L(t)dt $). The right panel shows the cumulative luminosity evolution. The purple, blue, cyan, green, orange and red line show the luminosity of the H$\alpha$ emission line, FUV band, NUV band, u band, g band, WISE-3 band, and UV+IR, respectively. Clearly, the H$\alpha$ emission line has the fastest reaction timescale ($50\%$ of the total luminosity is emitted in the first $\sim3$ Myr), followed by the FUV and other traces which test redder part of the spectrum. } 
    \label{fig:Levo}
\end{figure*}

As we have extensively discussed in the main text of the paper, the ability to differentiate different stochastic processes that govern star-formation is dependent on being able to measure the SFR with different indicators that probe different timescales. To be sensitive to a wider range of different PSDs, these response functions of the SFR indicators would ideally be well differentiated and cover a large range of timescales.  We show in Figure \ref{fig:Levo} the luminosity evolution, seen in different passbands, of a Simple Stellar Population (SSP). These are effective response functions, i.e., the observed SFH in any of these bands can be considered as a convolution of the intrinsic SFH with these luminosity evolutions. In the left panel, we see that all of the traces exhibit peak at early times ($\sim$ 5 Myr), followed by an exponential-like decline. The right panel shows the cumulative luminosity evolution, indicating more clearly the long term behaviour of the different indicators. H$\alpha$ responds very quickly to star-formation -- therefore, the measurement of the MS width in H$\alpha$ is very similar to the intrinsic MS width for almost the whole range of parameter space that we have considered, as seen in Figure \ref{fig:scatter_indicators}. As expected, the indicators corresponding to longer wavelengths tend to have slower responses, showing long ``memory'' of the previous SFH. The fact that all of these indicators are a mixture of the recent and long-term star-formation complicates matters and makes the interpretation of the measurements, such as width of MS, quite unintuitive, but they can be modelled through simulations (Figure \ref{fig:scatter_indicators}).\par

\section{Timescales of different SFR tracers} \label{sec:Timescales}

\begin{figure*}
    \centering
    \includegraphics[width=\textwidth]{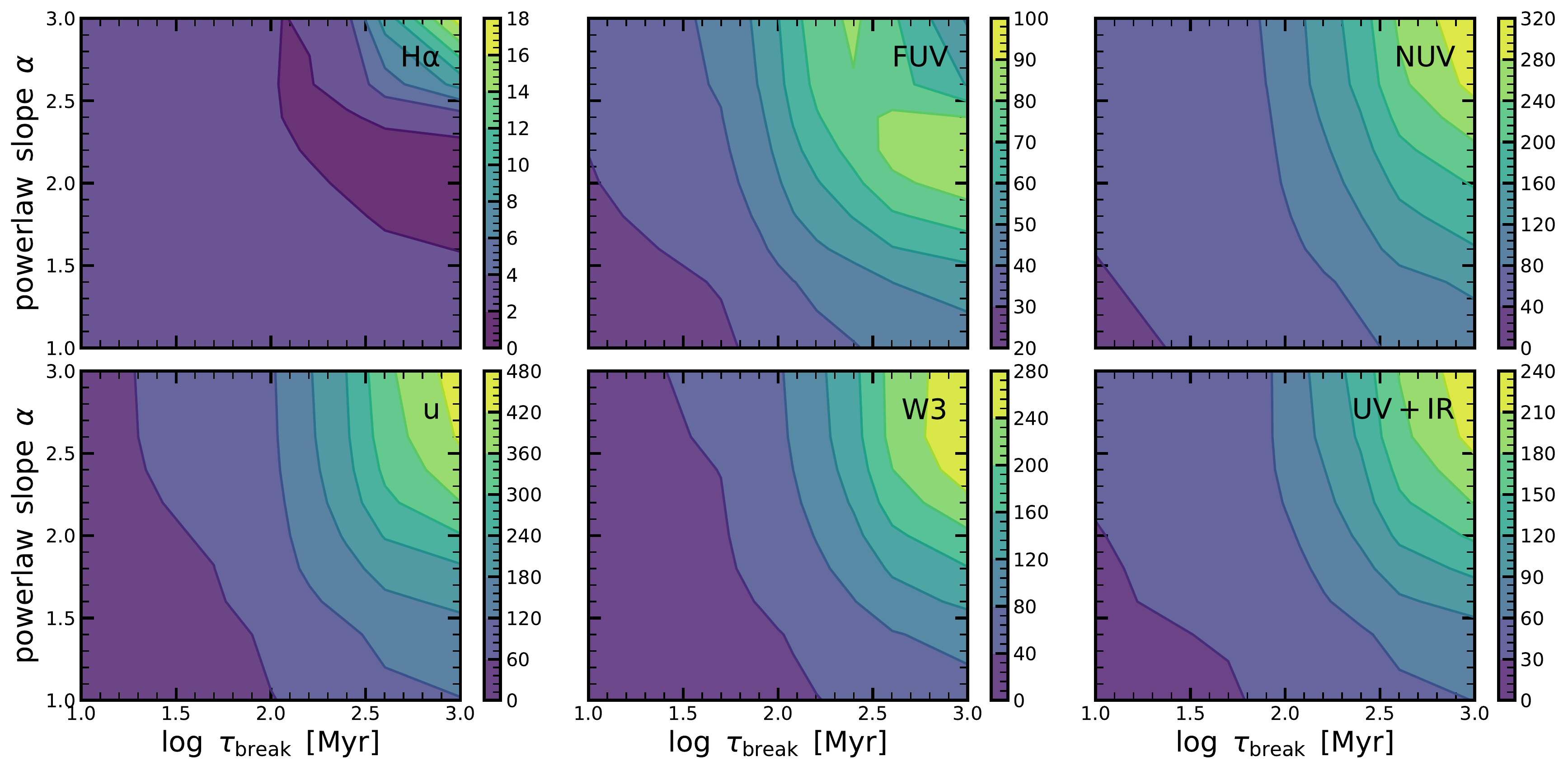}
    \caption{Effective timescale for different SFR tracers, as a function of parameters describing the underlying stochastic process ($\alpha$ and $\tau_{\rm break}$). Effective timescale is defined as a timescale over which the averaged SFH is most similar to actual SFR measured in a given indicator, and is indicated by the colour bar for each panel. This timescale depend not only on the indicator used, but also on the parameters of the stochastic process. To the first order, the burstier the SFH (see Section  \ref{sec:burst}), the shorter the timescale. Furthermore, the H$\alpha$ and FUV SFR indicators trace the shortest timescale.} 
    \label{fig:timescales}
\end{figure*}

We have shown in the main body of the manuscript how the key relations are simplified and well understood in the case where the SFR response function is a simple step function. On the other hand, in Appendix \ref{app:response_functions} we show that the real response functions have non-trivial time dependence and that they are clearly not well described by a step function. Crucially, the timescales that best approximate different SFR tracers actually depend on the property of the SFH itself. In practice, similar to the derivation of the luminosity-SFR calibration, one usually assumes a constant SFR and then determines the luminosity-weighted mean age of the stellar population \citep[e.g.,][]{kennicutt98, kennicutt12, madau14}. A more sophisticated approach is to consider more realistic SFHs from numerical and analytical simulations, with the caveat that the variability of the SFHs is simulation-dependent and it might bias the measurement \citep[e.g.,][]{tacchella18}. Already \citet{johnson13} showed -- based on empirical SFHs -- that fluctuations in the SFHs alone can cause factor of $\sim2$ variations in the UV luminosities relative to the assumption of a constant SFH over the past 100 Myr. Thanks to our framework of stochastic SFHs introduced in Section \ref{sec:Description}, we are able to generate and understand SFHs that have different variability properties according to a power spectral density.\par

Here we are looking to derive a singular timescale over which to average the realistic star-formation response to best approximate complex and realistic functions with a step function of a given duration. We will first briefly express our problem analytically and afterwards derive these timescales numerically for the different indicators that we used in Section \ref{sec:RealResponse} (H$\alpha$, FUV, NUV, u, W3 and UV+IR).\par

The statement about approximating realistic star-formation responses as a step function can be expressed as a search for the best $N_{\rm approx}$ to approximate Equation \eqref{masterSFRequation}:
\begin{equation} \label{eq:condition}
<\Delta(t_{i})>_{N_{\rm approx}}\cong  \log \sum^{\infty}_{t_{j}=1} 10^{\Delta (t_{i}-t_{j})} w_{\rm ind}(t_{j}),
\end{equation}
where 
\begin{equation} \label{eq:SFR_mean}
<\Delta(t_{i})>_{N_{\rm approx}}=\log \sum^{N_{\rm approx}}_{t_{j}=1} 10^{\Delta (t_{i}-t_{j})} .
\end{equation}
Comparing the expressions within the logarithms we see that the condition simplifies to the following equation for $N_{\rm approx}$:
\begin{equation}
\sum^{N_{\rm approx}}_{t_{j}=1} 10^{\Delta (t_{i}-t_{j})} =\sum^{\infty}_{t_{j}=0} 10^{\Delta (t_{i}-t_{j})} w_{\rm ind}(t_{j})
\end{equation}
From here we can see that in order to derive $N_{\rm approx}$ we have to consider both the properties of stochastic process, that govern time dependence of $\Delta(t_{i}-t_{j})$ term, and various indicators with their response function described with  $w_{\rm ind}(t_{j})$. In other words, as stated before, the timescale associated with each SFR traces depend on the previous SFH (in addition to the properties of each indicator).\par

In order to derive the timescales of the different SFR tracers numerically, we generate SFHs with different $\alpha$ and $\tau_{\rm break}$, as described in Section \ref{sec:Description}. We estimate the timescale of the SFR tracer by solving Equation \eqref{eq:condition} in a statistical sense, by minimizing the RMS difference of the SFR of the tracer and the intrinsic SFR averaged over different timescales. In other words, we search for $N_{\rm approx}$ that minimizes the variance of the quantity $\Delta_{\rm ind}(t_{i})-<\Delta (t_{i})>_{N_{\rm approx}}$.\par

In Figure \ref{fig:timescales}, we present the result of this analysis for H$\alpha$, FUV, NUV, u, W3 and UV+IR. As expected, H$\alpha$ corresponds to shortest timescale while UV+IR and u-band correspond to longer timescales (e.g., see Figure  \ref{fig:Levo}). We however note large variations across the parameter space of $\tau$ and $\alpha$, confirming our intuition about the importance of considering the underlying stochastic process on making a statement about timescales probed by different indicators. The differences are obviously larger for the longer indicators. Intuitively, when the indicator is very quick, as is the case for H$\alpha$ or FUV, effect of ``mixing in'' of the underlying process is reduced and changes of $\tau_{\rm break}$ and $\alpha$ that affect longer timescale are not probed in this analysis. \par

For the set of parameters that we have deduced in Section \ref{sec:Observational}, $\alpha=2$ and $\tau_{\rm break}=178^{+104}_{-66}$  Myr, we find that the effective times of the indicators are $t_{\rm eff} (\mbox{H}\alpha)=2.0_{-0.1}^{+0.4}~\mathrm{Myr}$, $t_{\rm eff} (\mbox{FUV})=63_{-1}^{+2}~\mathrm{Myr}$ , $t_{\rm eff} (\mbox{NUV})=119_{-2}^{+8}~\mathrm{Myr}$ , $t_{\rm eff} (\mbox{u-band})=164_{-16}^{+15}~\mathrm{Myr}$, $t_{\rm eff} (\mbox{W3})=96_{-14}^{+7}~\mathrm{Myr}$,  and $t_{\rm eff} (\mbox{UV+IR})=84_{-5}^{+6}~\mathrm{Myr}$. Of course, these estimates suffer from the same caveats that we have discussed in Sections \ref{sec:Observational} and just therefore only be taken as a general guidance, rather than strict quantitative measurement. \par 
\bsp	
\label{lastpage}
\end{document}